\documentclass[conference]{IEEEtran}
\IEEEoverridecommandlockouts
\usepackage{cite}
\usepackage{amsmath,amssymb,amsfonts}
\usepackage{algorithmic}
\usepackage{graphicx}
\usepackage{textcomp}
\usepackage{xcolor}
\usepackage{float}
\def\BibTeX{{\rm B\kern-.05em{\sc i\kern-.025em b}\kern-.08em
    T\kern-.1667em\lower.7ex\hbox{E}\kern-.125emX}}
\usepackage[spaces,hyphens]{xurl} 
\usepackage[colorlinks,allcolors=blue]{hyperref}
\usepackage{siunitx}
\usepackage{subcaption}
\usepackage{array, tabularx, longtable, ltablex, booktabs}

\newcommand{\tb}[1]{\textbf{#1}}
\begin{document}

\title{Death By A Thousand COTS: Disrupting Satellite Communications using Low Earth Orbit Constellations\\
}


\author{\IEEEauthorblockN{Frederick Rawlins}
\IEEEauthorblockA{
\textit{University of Oxford}\\
}
\and
\IEEEauthorblockN{Richard Baker}
\IEEEauthorblockA{
\textit{University of Oxford}\\
}
\and
\IEEEauthorblockN{Ivan Martinovic}
\IEEEauthorblockA{
\textit{University of Oxford}\\
}
}

\maketitle

\begin{abstract}
Satellites in Geostationary Orbit (GEO) provide a number of commercial, government, and military services around the world, offering everything from surveillance and monitoring to video calls and internet access. However a dramatic lowering of the cost-per-kilogram to space has led to a recent explosion in real and planned constellations in Low Earth Orbit (LEO) of smaller satellites. 

These constellations are managed remotely and it is important to consider a scenario in which an attacker gains control over the constituent satellites. In this paper we aim to understand what damage this attacker could cause, using the satellites to generate interference. 


To ground our analysis, we simulate a number of existing and planned LEO constellations against an example GEO constellation, and evaluate the relative effectiveness of each. Our model shows that with conservative power estimates, both current and planned constellations could disrupt GEO satellite services at every groundstation considered, with effectiveness varying considerably between locations. 

We analyse different patterns of interference, how they reflect the structures of the constellations creating them, and how effective they might be against a number of legitimate services. We found that real-time usage (e.g. calls, streaming) would be most affected, with 3 constellation designs able to generate thousands of outages of 30 seconds or longer over the course of the day across all groundstations.

\end{abstract}



\section{Introduction}

\subsection{Motivation}

Space infrastructure underpins an array of services that the world relies upon, including GNSS for navigation and timing, broadcast media and internet access in remote areas. Historically, the cost and complexity of developing and deploying space infrastructure was so vast that it was accessible only to the most well-funded government and commercial entities. However, a `New Space' revolution is underway, bringing orbital operations within the reach of a far larger number of entities, enabled by the development of flexible, commodity hardware; multi-tenanted launches; and an ecosystem of support services that have come about together \cite{NewSpace}.

Noting this vast expansion in the quantity of space infrastructure, emergence of new services (presenting more attack surface), and the newfound ease with which more participants can launch their own satellites, there is a novel potential for Downlink Interference Attacks (DIAs) by current constellations and those planned for the future. Despite the substantial disparity in transmission power between the attacker and victim satellites, the difference in orbital distance and the far greater number of potential attacking satellites make an attack such as this important to consider. 

\begin{figure}[t]
    \centering
    \includegraphics[width=9cm]{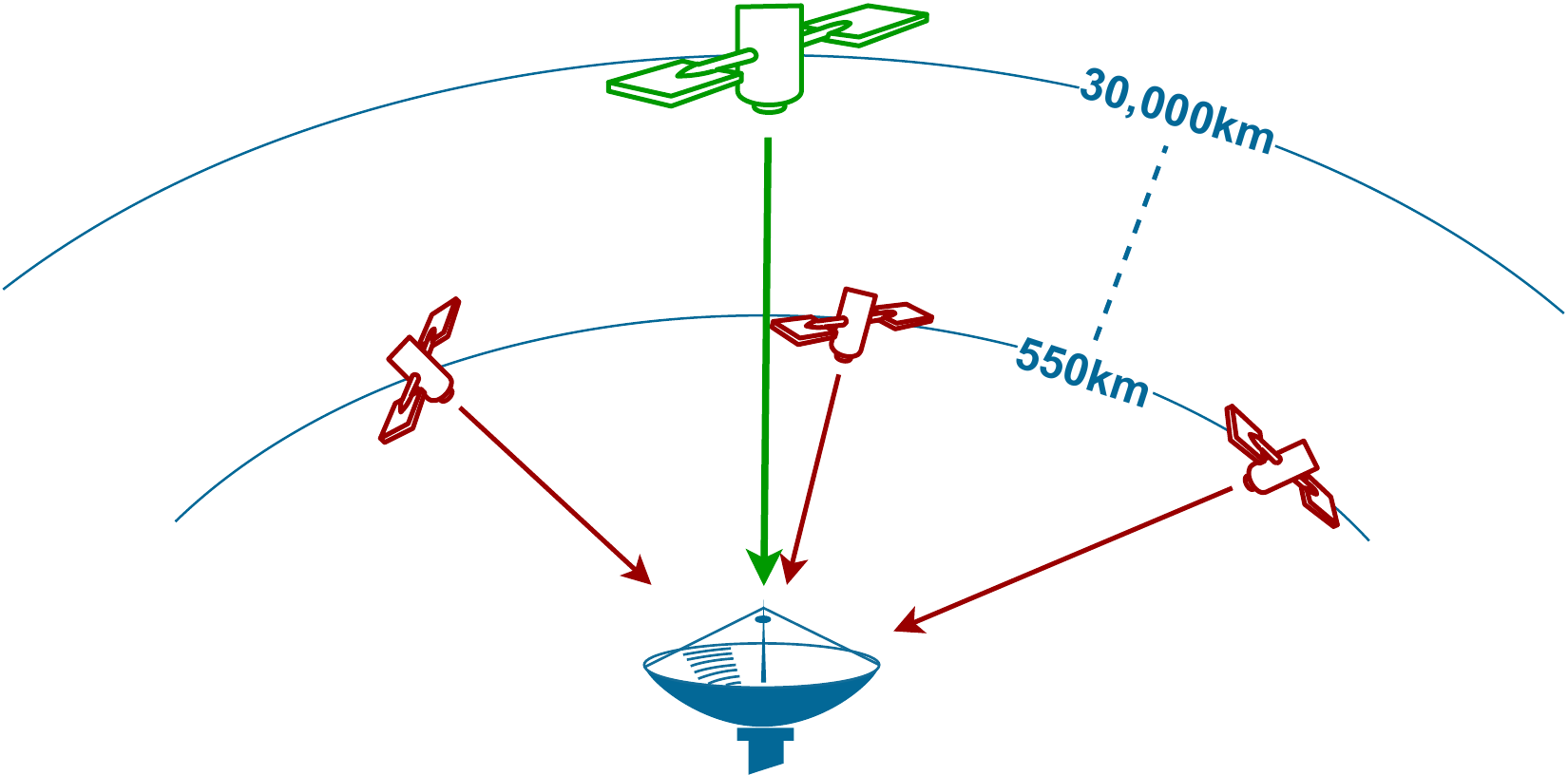}
	\caption{An illustration of the attack setup. Three LEO satellites with signals reaching the victim receiver, interfering with the legitimate signal (centre).}
	\label{fig:IntroSatelliteImage}
\end{figure}

For years, terrestrial Distributed Denial of Service (DDoS) attacks have mostly used `botnets' composed of thousands of compromised computers to disrupt services across the internet \cite{botnet}. An attack of this scale may seem disproportionate for space, but there are a number of multi-thousand satellite constellations planned.
It is therefore worth examining whether these constellations might represent the same threat. They would operate on slightly different principles, relying on the fact that with a sufficiently dense constellation, there will always be a satellite close to the line between the victim satellite and groundstation, where interference would be strongest. However, the same attack vector exists: thousands of near identical systems that cannot be air-gapped, and could potentially be compromised from almost anywhere on Earth.

Our paper seeks to explore the following research questions:

\begin{enumerate}
    \item How much interference can different LEO constellations generate?
    \item How is interference affected by the location of the receiver?
    \item Which of the current or planned constellations would be most effective (and therefore potentially more appealing as a target of malware)?
    \item How are different legitimate uses of GEO satellites affected?
\end{enumerate}


%
%
%

\subsection{The Attack}

\begin{figure}[ht]
    \centering
    \includegraphics[width=9cm]{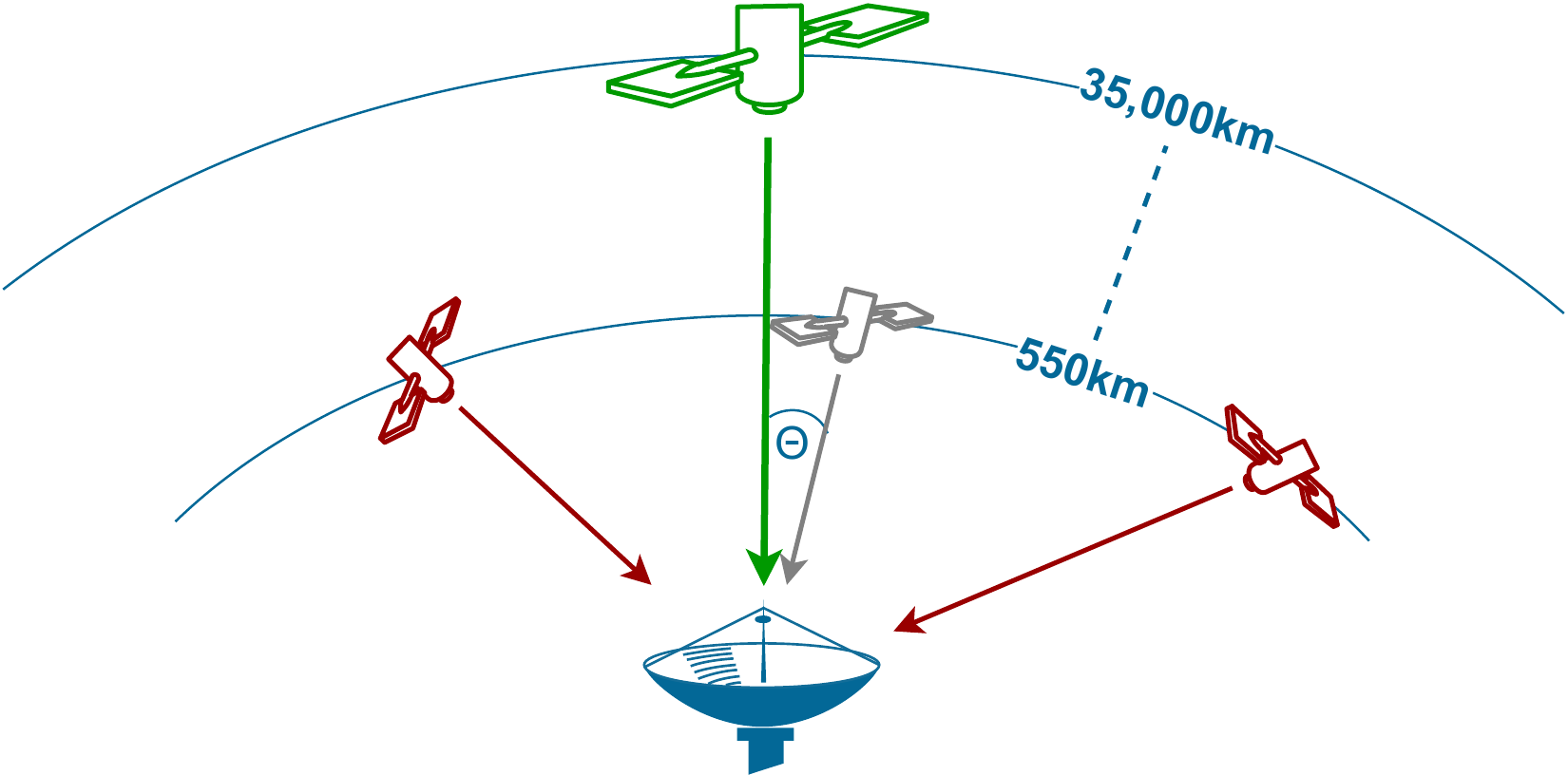}
	\caption{Current interference avoidance mechanisms based on selectively disabling communications with low angular separation ($\theta < \epsilon$ for some threshold $\epsilon$).}
	\label{fig:AttackCase}
\end{figure}

It is recognised that interference between satellites operating at the same frequencies can lead to degraded communications \cite{SignificantRulings}. As such, modern LEO constellations are required by regulators like the Federal Communications Commission (FCC) to address how they plan to avoid interference with existing infrastructure. As evidenced from the filings themselves, they take a good-faith, proactive approach to this by ``ensuring that there is the necessary amount of angular separation'' between themselves and potential victim satellites when targeting groundstations \cite[p. 27]{OneWebTechnical}. This works because the receiving antenna will have much greater reception (gain) at the angle directly towards its intended satellite. This interference-avoidance therefore relies on each component of the constellation continually monitoring its own position relative to a number of other satellites to avoid broadcasting in case of insufficient angular separation as seen in Fig.~\ref{fig:AttackCase}. 

In our case, the attacker either does not implement this measure, or overrides it, and we see the re-enabling of the centre satellite as shown in Fig.~\ref{fig:IntroSatelliteImage}. 


\subsection{Contributions}

\begin{itemize}
    \item We establish a simulation framework for Downlink Interference Attacks.
    \item We evaluate current and planned constellations' effectiveness at jamming the Inmarsat Global Xpress system in the Ku/Ka bands.
    \item We find notably that some existing constellation would be sufficient to systematically jam Inmarsat's service around the world. 
    \item We discuss the limitations of CubeSats in the communications space, and their potential for disrupting GNSS services.
\end{itemize}

\section{Background}

\subsection{SDRs in Space}

SDRs have dramatically lowered the barrier to entry for satellite operators, and increased flexibility. The ability to change frequency, modulation, encoding schemes, and processing on the fly enables a much more dynamic and adaptable system.

Licenses for space operation give strict frequency bands and operating parameters for satellites. As identified in \cite{SameBoat}, a satellite developer must provide a transmitter survey and potentially verification of emission characteristics, however these mostly rely on a benign satellite operator providing trustworthy data \cite{CubeSat101, RangeSafety}. The adversarial analysis done by Pavur et al. shows that physical changes are limited by size and power, but software is extremely difficult to completely externally verify without a detailed code review, which is currently neither conducted nor required. 

This means satellites verified for one mode of operation on the ground can change once in orbit. They have the physical capability through the use of SDRs and wide-band antennas to broadcast at frequencies beyond their license, and due to the congested nature of frequency band allocation, this possible band of operation will almost certainly overlap with other satellites.

\begin{figure}[!t]
    \includegraphics[width=8cm]{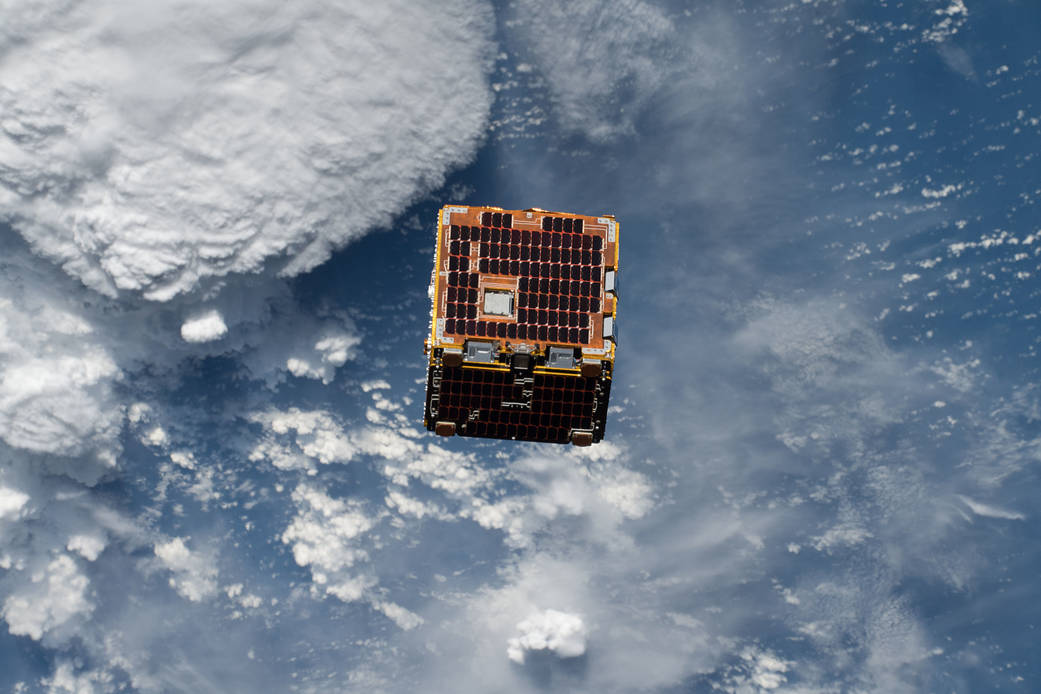}
	\caption{CubeSat in Orbit - \textit{NASA}}
	\label{fig:marvin}
\end{figure}

\begin{figure*}[t]
    \begin{subfigure}{0.33\textwidth}
        \includegraphics[width=\textwidth]{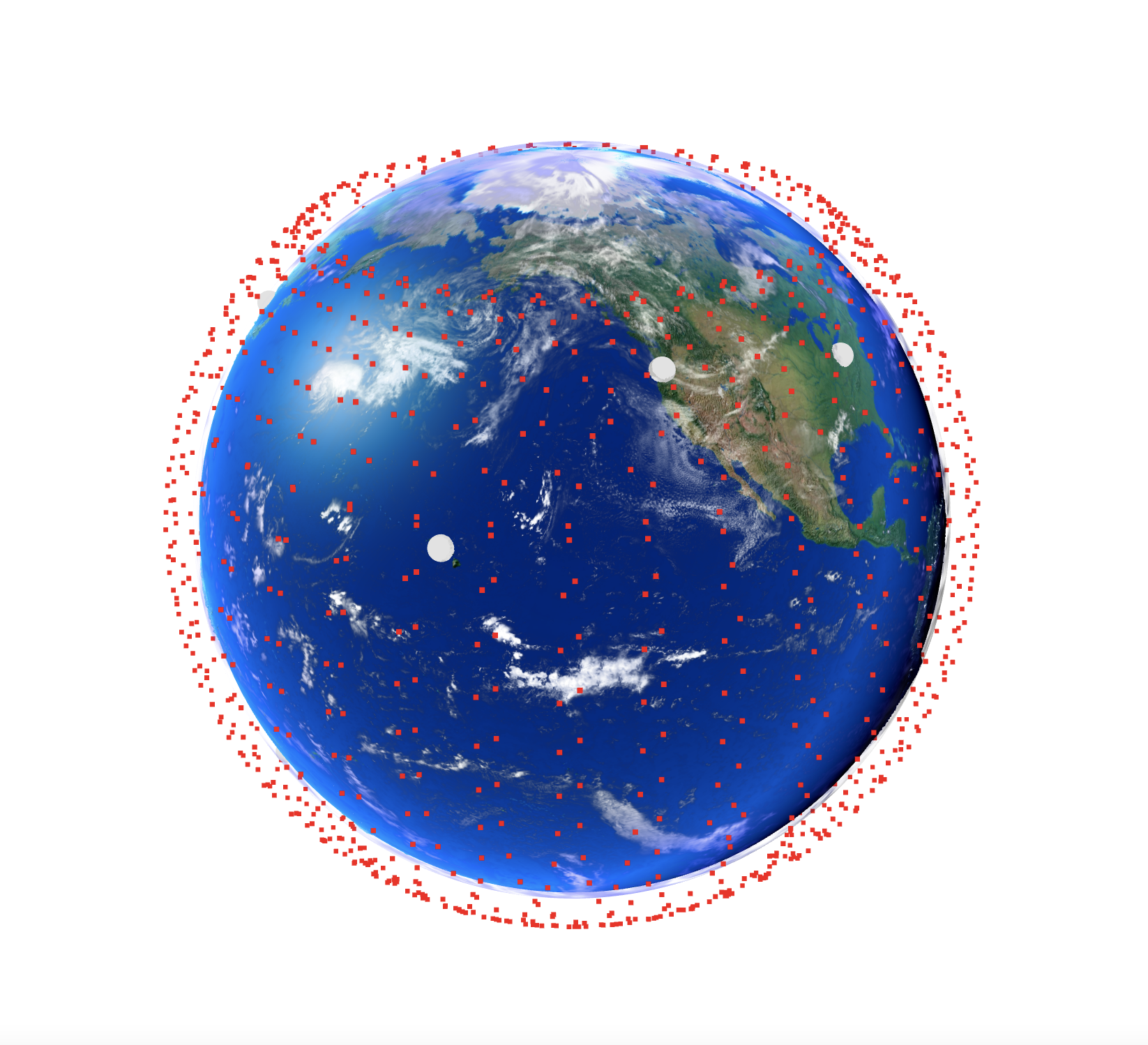}
        \caption{Starlink}
    \end{subfigure}
    \begin{subfigure}{0.33\textwidth}
        \includegraphics[width=\textwidth]{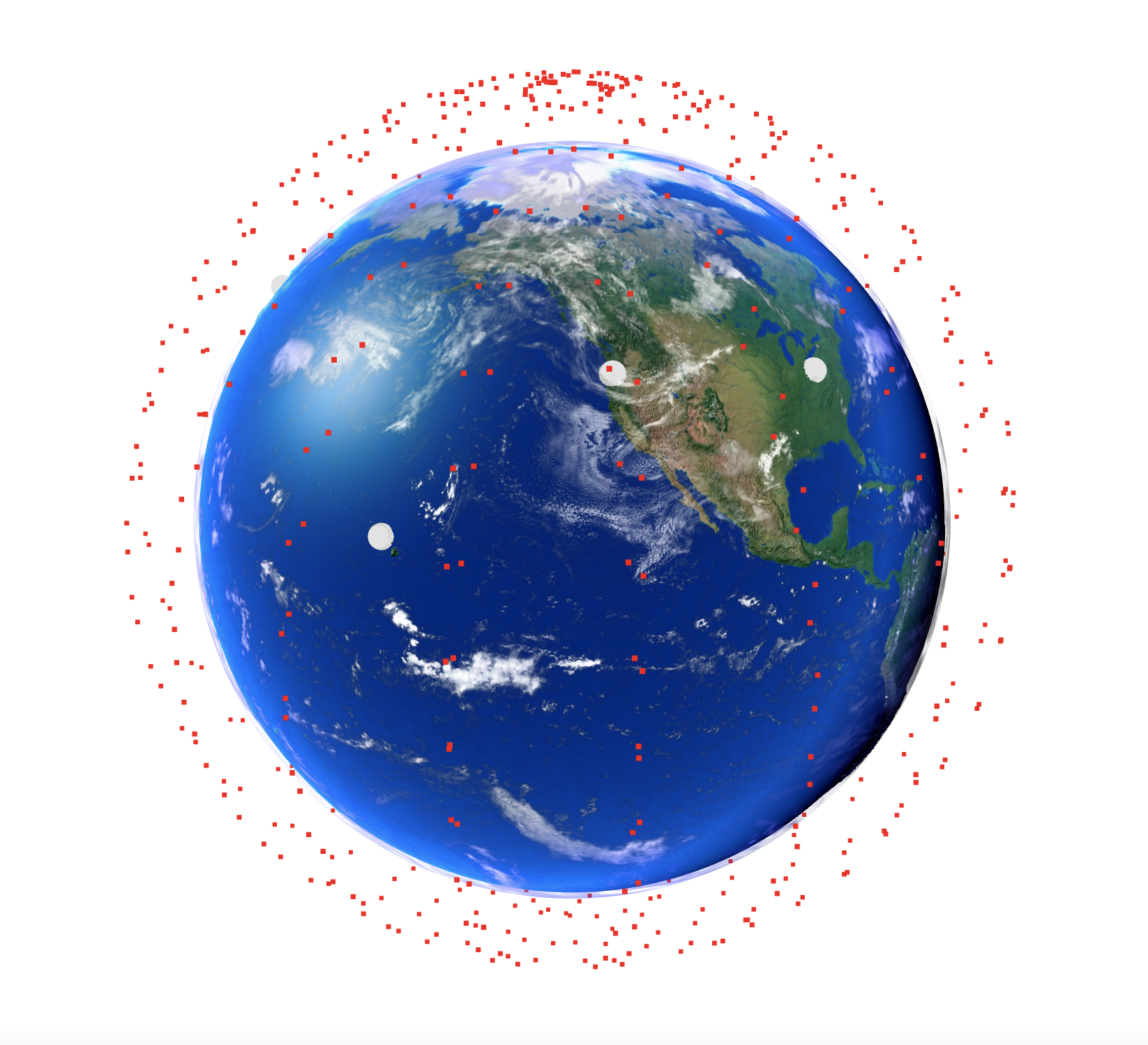}
        \caption{OneWeb}
    \end{subfigure}
    \begin{subfigure}{0.33\textwidth}
        \includegraphics[width=\textwidth]{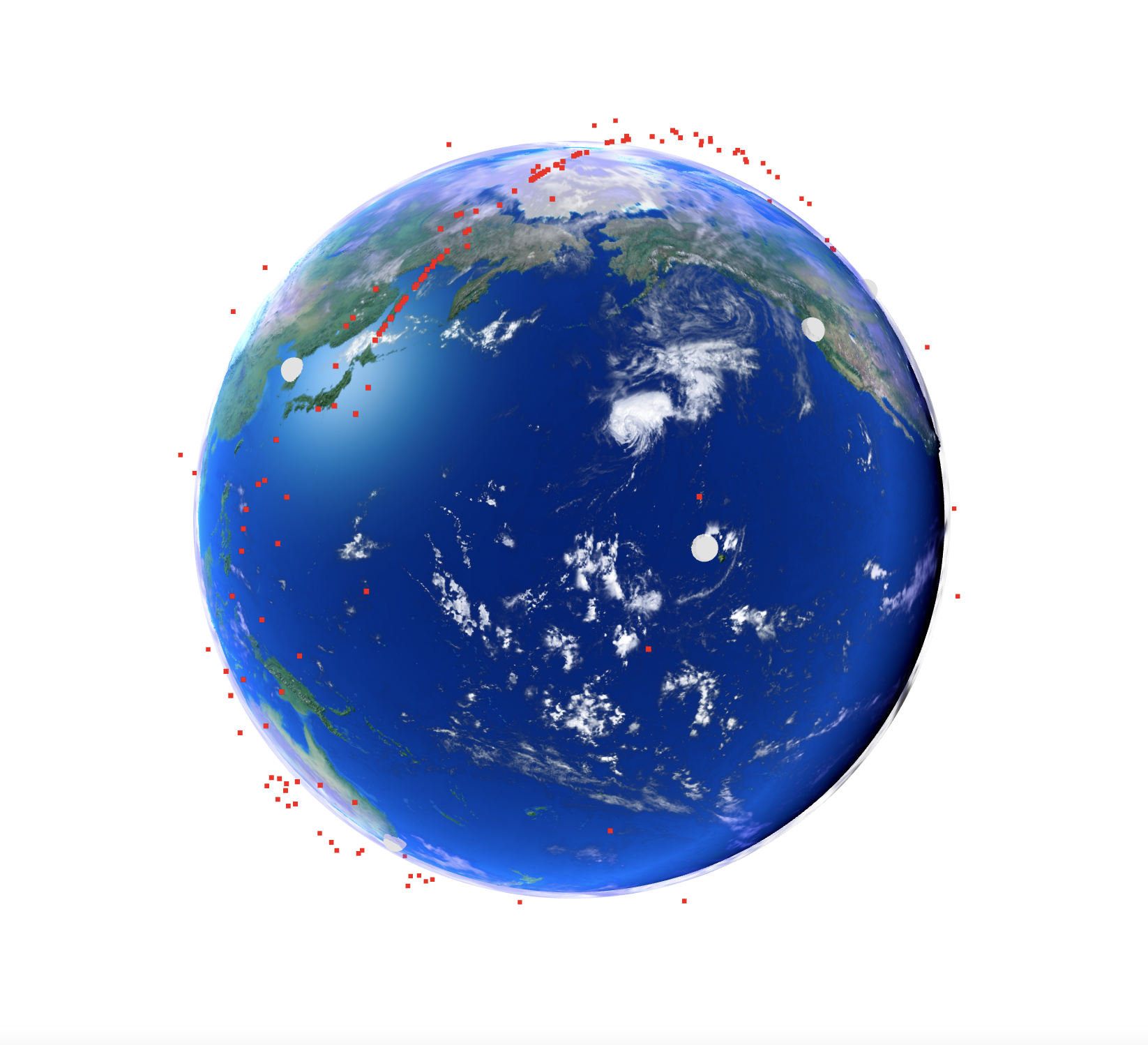}
        \caption{Planet Labs's Dove}
    \end{subfigure}
    \caption{Three of the constellations used, each with a different structure.}
\end{figure*}

\subsection{CubeSats}

Satellite architecture is one of the key changes in the shift towards low-cost `small sats' \cite{ModernSmallSats}. This has been dubbed the era of `New Space' \cite{NewSpace}, and the clear differences in both technology and attitude present a host of potential vulnerabilities, while also improving on some specific areas of security.

`Old Space' satellites were almost all built by governments or large defence/communication companies. Every satellite was custom-built for its mission over a span of years, and excessively tested on the ground to ensure its success and reliability in space. Not only would each satellite have been extremely expensive, as it was using proprietary hardware and software, its trip to space would have been too, coming in at \$54,500/kg\cite{LaunchCost}. At an average mass of 3,000kg\footnote{Exact value: 3,002kg. Calculated as the average of all launch masses of satellites currently in orbit greater than 500kg.\textit{Database:} \url{https://www.ucsusa.org/resources/satellite-database}.}, that gives a launch cost of \$163 million. These satellites had very little focus on security, and extensive interception of signals from the Iridium Constellation has already been well documented\cite{Iridium}. Iridium's reliance on security-through-obscurity was shared by many other manufacturers and operators of similar size satellites, and with the advent of SDRs these satellites are rapidly becoming more accessible to malicious third parties.

`New Space' satellites on the other hand make use of commercial-off-the-shelf (COTS) hardware for the most part, drastically reducing costs and development time. This is largely in part due to the CubeSat Design Specification, which has presented a modular standard for others to build upon. Kits are available online providing a majority of the CubeSat components pre-built for as little as \$7,000\cite{ESat}. Even non-CubeSats, like the satellites in the Starlink constellation, are using COTS hardware alongside their own proprietary hardware.

In general these CubeSats are significantly less powerful than traditional `Old Space' satellites, since they focus on being small, light, and easy to build. However this cheap approach is reflected not only in their physical hardware and software, but also in their orbit. Many larger satellites expend a great deal of energy, propellant, and money to achieve a Geo-stationary Orbit' (GEO) (roughly 35,000km away). This is the perfect distance from Earth where a satellite's orbit matches the rotation of the Earth underneath it, allowing it to remain `fixed' in the sky. and able to view a large percentage of the ground (sometimes up to 1/3$^{\text{rd}}$ of Earth's surface area). This is also used for imaging and defence satellites to enable them to focus on a specific place the entire time. CubeSats take a much cheaper approach. They exist in Low Earth Orbit (LEO) instead, (\textasciitilde{}550 km), often because a rocket can carry a large number of small satellites at once to this lower orbit (sometimes 100+\cite{SXT3}), reducing the price for each customer. Another key advantage of LEO lies in the shorter signal path to Earth and back. What the satellites lose in coverage, they dramatically gain in signal strength. This is understandably important when using COTS SDRs and antennas, since much more powerful communications systems would be required in GEO, or on Earth to compensate. The final advantage, and the most crucial for satellites like Starlink, is the lower latency. In order to achieve the responsiveness required for real time video calls, gaming, and streaming, the latency must be at least of the same order of magnitude as terrestrial broadband (20 ms). It's clear then that GEO, where the round trip for light is 200 ms not including any processing at either end, would never be feasible, whereas Starlink in its orbit of roughly 550 km has been measured to have a latency of 45 ms.

As established however, each satellite in LEO can see much less of the Earth at any given point. Moreover, it is only above that point for a matter of minutes before its rapid orbit takes it over the horizon. If global coverage is desired, then hundreds or thousands of satellites are required, which is the basis for constellations.

\newcolumntype{x}[1]{>{\centering\arraybackslash\hspace{0pt}}p{#1}}

\begin{table*}[!t]
\footnotesize
\centering
\begin{tabular}{
p{1.4cm}
x{1.4cm}
x{1.4cm}
x{1.4cm}
x{1.4cm}
x{1.4cm}
x{1.4cm}
x{1.4cm}
x{1.4cm}
x{1.4cm}
x{1.4cm}
x{1.4cm}}
\toprule
\tb{Attack}  & \tb{Military} & \tb{Intelligence Agency} & \tb{Corporate Insider} & \tb{Hardware Supplier} & \tb{APT} & \tb{Organised Crime} & \tb{Terrorist Group} & \tb{Individual Hacker} & \tb{Activist Group} \\
\midrule
Constellation Launch & 
$\textcolor{teal}{\checkmark}$ & $\textcolor{teal}{\checkmark}$ & $\mathrm{x}$ & $\mathrm{x}$ & $\mathrm{x}$ & $\textcolor{orange}{i}$ & $\textcolor{orange}{i}$ & $\mathrm{x}$ & $\textcolor{orange}{i}$ \\
\midrule
Pre-Launch Tampering &
$\textcolor{teal}{\checkmark}$ & $\textcolor{teal}{\checkmark}$ & $\textcolor{teal}{\checkmark}$ & $\textcolor{teal}{\checkmark}$ & $\textcolor{orange}{i}$ & $\textcolor{orange}{i}$ & $\textcolor{orange}{i}$ & $\textcolor{orange}{i}$ & $\textcolor{orange}{i}$ \\
\midrule
In-Space Tampering &
$\textcolor{teal}{\checkmark}$ & $\textcolor{teal}{\checkmark}$ & $\textcolor{teal}{\checkmark}$ & $\mathrm{x}$ & $\textcolor{teal}{\checkmark}$ & $\mathrm{x}$ & $\textcolor{orange}{i}$ & $\textcolor{orange}{i}$ & $\textcolor{orange}{i}$ \\
\midrule
Distributed Jamming &
$\textcolor{teal}{\checkmark}$ & $\textcolor{teal}{\checkmark}$ & $\mathrm{x}$ & $\mathrm{x}$ & $\textcolor{teal}{\checkmark}$ & $\mathrm{x}$ & $\mathrm{x}$ & $\textcolor{orange}{i}$ & $\textcolor{orange}{i}$ \\
\bottomrule
\\
\multicolumn{10}{p{\textwidth}}{
    Key: $\textcolor{teal}{\checkmark}$ - Attacker is likely both capable of executing the attack and motivated to do so. c - Attacker is likely capable, but the vulnerability doesn't align with motivations. $\mathbf{i}$ - Attacker is likely interested in the attack, but has limited capacity to execute it. $\mathbf{x}$ - Attacker is likely neither interested in nor capable of executing the attack. Note: There may be crossover between categories, such as an insider threat sponsored by an intelligence agency. This matrix is intended as a demonstrative summary of the four major stages of the attack, with one of the first three required for execution of the fourth.
}
\end{tabular}
\caption{Threat Matrix}
\label{tab:threatMatrix}
\end{table*}

\subsection{Related Work}

The main body of intersatellite interference research, small though it is, focuses on ensuring that the good faith solutions proposed work to keep interference at acceptable levels, and the side effects on both systems. The effectiveness of angular separation and beam control in reducing interference have been studied with a view to understanding the cost to LEO coverage that it imposes \cite{ExclusiveAngle,OptimalBeam}, and steps have been taken to find analytical alternatives to orbital simulation for efficient interference calculation \cite{Cofrequency}. 

These studies however do not take an adversarial perspective or consider a threat model, and so are not able to provide an answer for a number of our questions surrounding the potential capabilities of different actors. They also lack a comparative component considering often a single constellation at a time, making it much more difficult to generalise the results or find which constellations might be more appealing for takeover/creation.

\textit{``Should We Worry About Interference in Emerging Dense NGSO Satellite Constellations?''} \cite{ShouldWeWorry} studies a number of constellations and their interference effects, but does so with a Monte Carlo approach to satellite positioning. This Monte Carlo approach provides a more even distribution of relative positions, but does not allow us to see how interference changes over time. This is crucial when considering how different interference patters affect the user's ability to interact with the system, as explored in Section \ref{sec:UsageAnalysis}.

\section{Threat Model}

Building upon the work done by Pavur et. al to create a template for understanding space-based threat models \cite{SOKpavur}, in Table~\ref{tab:threatMatrix} we present a threat matrix covering major threat actors and capabilities.


The victim considered is a legitimate user of the GEO satellite(s) for communication or navigation purposes. This might be downloading files, using it for voice/video calls, receiving location signals, or browsing the internet. 


An attacker in this scenario has either the capability to take control of a substantial portion of an existing constellation, be it through malicious cyber attack or state-level take-over, or funding to place a legitimate constellation into orbit. Their aim is to stop a user on the ground from being able to extract a legitimate GEO signal from the noisy signal they record, rendering them unable to receive data.

The specific details of a cyber attack on a constellation are beyond the scope of this paper, but with growing focus on this potential threat \cite{hackasat, aerospaceVillage}, and increasing use of COTS components/software familiar to existing groups \cite{TheHardwareInSpace}, it is a realistic proposal to consider.

The number of entities with the resources to produce, launch, and maintain a constellation (even of CubeSats) on the other hand is relatively small. It requires a large upfront investment in both money and time, covering development, launch, and orbit raising. Commercial Off The Shelf components have reduced this investment, but it is still non-trivial. 

A few well-funded university research groups might be able to launch 1-10 CubeSats on a ride-share \cite{virginia}. A consortium of universities and labs could extend this to a few dozen\cite{qb50}. From the company perspective, Swarm Technologies were able to put 120 0.25U CubeSats into orbit\cite{swarm}, but barring the major constellations such as Starlink, OneWeb, or Iridium, almost every company with orbital assets has launched fewer than 100 satellites\cite{celestrak}.




\section{Radio Modeling}
\label{sec:Model}

To estimate interference we need to understand how the signals interact with each other at the ground. The signal strength depends on the frequency of signal, the distance between the transmitter and the receiver, the power and gain of transmitter and receiver, the angle between them, and signal loss to atmosphere as seen in Eq.~\ref{eq:transmission}. We simulate the satellite's orbital propagation (Section~\ref{sec:Simulation}) and at each time step, with the distances and angles recorded, we can use Eq.~\ref{eq:SINR} to determine the received power from each satellite. This allows us to calculate the resulting Signal to Interference plus Noise Ratio (SINR), by taking the strongest legitimate signal received from the GEO satellites, and the sum of squares from interfering signals combined with receiver noise. 

SINR gives us an indication of `signal clarity', and is used to determine whether a groundstation is likely to be able to receive and correctly decode the legitimate signal.

\subsection{Transmission Equations}
The model itself is a basic representation of RF signals, written as the following:
\begin{equation}
    \label{eq:transmission}
    r_p = \textit{EIRP} + r_g - L - A
\end{equation}
\begin{equation}
    L = 20 \cdot \log \left( \dfrac{4 \pi d f}{c}  \right)
\end{equation}

For the components of the ratio, power is measured in watts (W) rather than decibel watts (dBW) as above, though the final SINR is in decibel watts (dBW).
\begin{align}
    \label{eq:SINR}
    SINR    &= \dfrac{S}{(N + I)} \\
            &= \dfrac{\max(C_i)}{N + \sqrt{\sum_{j=1}^n I_j^2}}
\end{align}

\begin{align*}
    \textit{EIRP} &= \text{Effective Isotropic Radiated Power (dBW)} \\
    r_p &= \text{Received Power (dBW)} \\
    r_g &= \text{Receiver Antenna Gain (dBW)} \\
    L &= \text{Free Space Path Loss (dBW)} \\
    A &= \text{Atmospheric Attenuation (dBW)} \\
    d &= \text{Distance (m)} \\
    f &= \text{Frequency (Hz)} \\
    b &= \text{Bandwidth (Hz)} \\
    c &= \text{Speed of Light (ms$^{-1}$)}
\end{align*}

\subsection{Receiver Equations}

The receiver is modeled according to the European Radiocommunications Committee's (ERC) report on analysis of Inmarsat receivers \cite{ERC} (with $G_{\max} = 44$ \cite{gxAntenna}) which for 19.2 GHz (Inmarsat's downlink frequency) gives:
\begin{align}
    G(\phi)=
    \begin{cases}
        G_{\max }-\left(\frac{D}{\lambda} \cdot \frac{\varphi}{20}\right)^{2} & \text { for } 0<\varphi<\varphi_{11} \\ 
        G_{1}                                                           & \text { for } \varphi_{m} < \varphi <100 \frac{\lambda}{D} \\ 
        52-10 \log \frac{D}{\lambda}-25 \log \varphi                    & \text { for } 100 \frac{\lambda}{D} \leq \varphi < 48 \\
        -10                                                             & \text { for } \varphi > 48
    \end{cases}
\end{align}



\begin{figure}[ht]
\begin{center}
\includegraphics[scale=0.5]{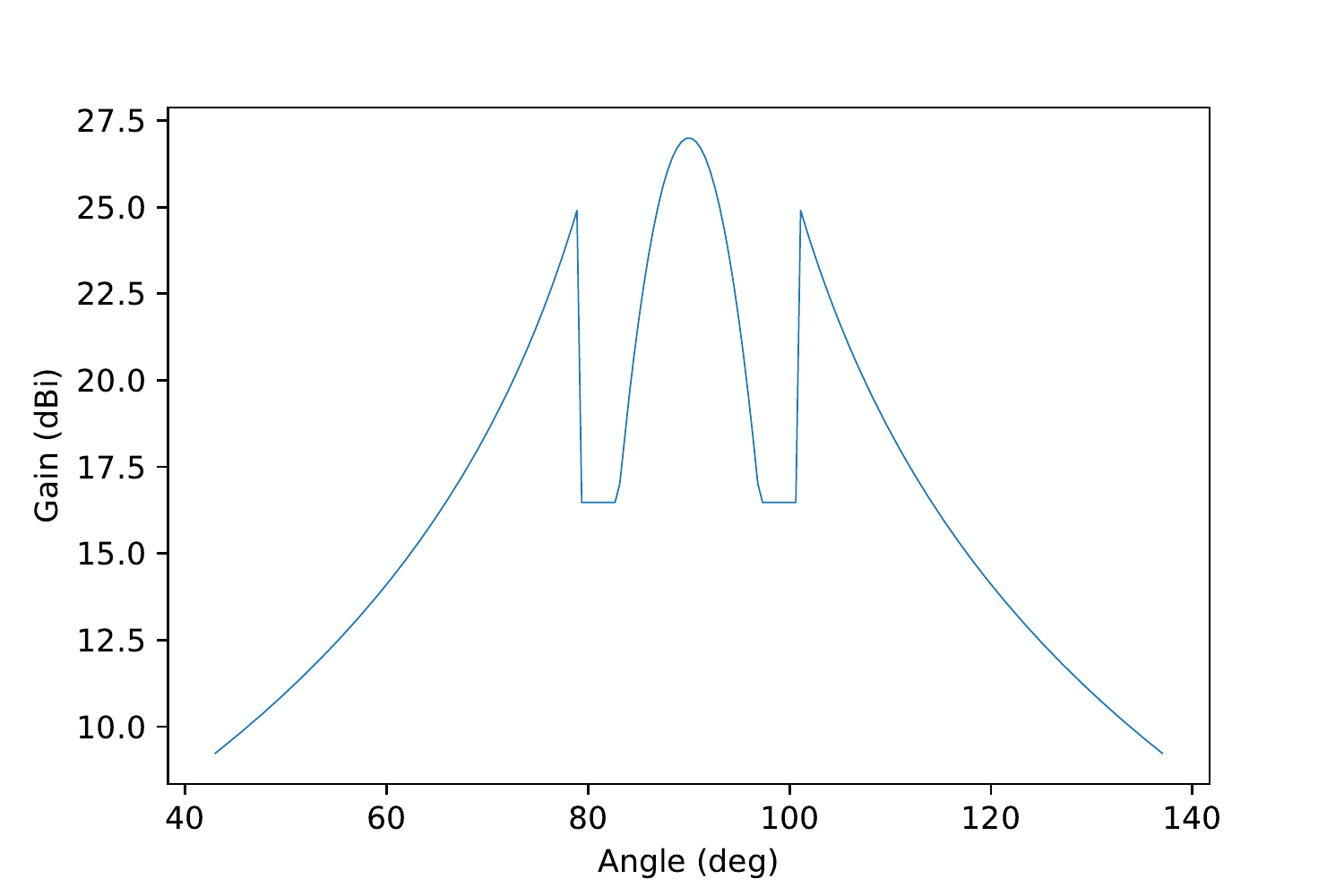}
\caption{The gain of the receiver modeled, where 90\si{\degree} is directly facing the GEO satellite, and LEO satellites will have a lower gain based on their angle to the receiver.}
\label{fig:Gain}
\end{center}
\end{figure}

This is adapted slightly since the values coming out of the simulation have $90\si{\degree}$ as directly overhead, so we end up with the graph in Fig.~\ref{fig:Gain}.

\subsection{Values}

The following values were used throughout:
\begin{align*}
    A &= \text{0.35 dBW} \\
    t &= \text{290 k} \\
    f &= \text{19.2 GHz} \\
    b &= \text{250 MHz}
\end{align*}
$A$ approximating atmospheric attenuation \cite{kymeta}, $t$ approximating room temperature for the receiver, $f$ set as the Inmarsat downlink frequency (still covered by the Ka and Ku Starlink transmitters), and $b$ for the bandwidth of the signal \cite{BoeingArctic}.

Rain attenuation is not modeled, but assuming clear skies for the simulation, this is negligible \cite{kymeta}.


\section{Evaluation}
\label{sec:Eval}

\subsection{Simulation}
\label{sec:Simulation}

\subsubsection{Simulator Details}
The satellite simulations were done in FreeFlyer, an orbital propagation program designed to simulate mutliple groundstations, satellites, and celestial bodies over a period of time. This takes known satellite positions and orbits, and steps through to calculate their position over time. A number of techniques are available to do this efficiently, though we use FreeFlyer's default  for simplicity \footnote{Runge Kutta 8(9) Integrator, with a fixed step size of 300s, and a Relative Error Tolerance of 1e-9, with Norad SGP4 for TLEs.}.

Atmospheric drag is included in the simulation, though not relevant for the timescales involved.

\begin{figure}
    \centering
    \includegraphics[width=220px]{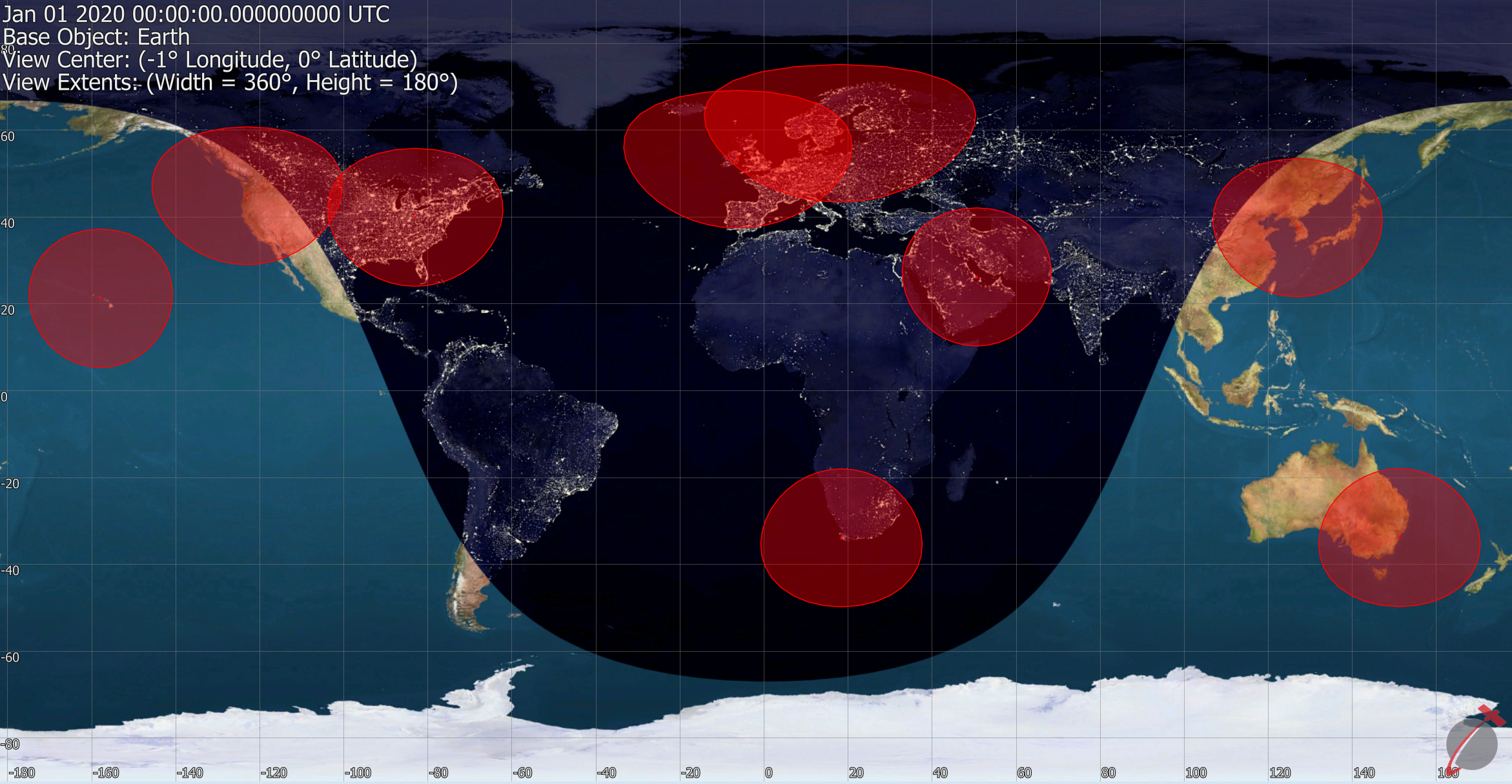}
    \caption{AWS Groundstation locations used as reference for receiver.}
    \label{fig:groundstations}
\end{figure}

For picking a relevant and varied group of Ground Station locations, we chose those listed as AWS Ground Stations \footnote{\url{https://aws.amazon.com/ground-station/locations/}} as show in Fig.~\ref{fig:groundstations}. These are 10 stations around the world, and as the Ground-Station-As-A-Service industry grows, may represent a substantial part of satellite-ground communications. While fairly evenly distributed longitudinally, they are focused mostly in the Northern Hemisphere. 

As listed in Section \ref{sec:Model}, we are using an Inmarsat specific antenna pattern, so the use of AWS groundstations is just for representative location rather than hardware.


\begin{table*}[t]
    \footnotesize
    \centering
    \begin{tabular*}{\textwidth}{
    p{4.4cm}
    x{2.2cm}
    x{2.2cm}
    x{2.2cm}
    x{2.2cm}
    x{2.2cm}}
    \toprule
    \tb{Constellation}  & \tb{Status} & \tb{Number} & \tb{Satellite} & \tb{EIRP (dBW)} & \tb{Orbit (km)}  \\
    \midrule
    Inmarsat Global Xpress \tb{(Victim)} & Exists & 150 & Global Xpress & 80 & 35,000 \\
    \midrule
    PlanetLabs Dove Constellation & Exists & 150 & Dove & 15 & 530 \\
    \midrule
    CubeSat Walker Constellation & Hypothetical & 396 & CubeSat & 6 & 550 \\
    \midrule
    Starlink First Group & Exists & 1,584 & Starlink & 45 & 550 \\
    \midrule
    Starlink Phase 1 & Planned & 4,408 & Starlink & 45 & 550 \\
    \midrule
    One Web Phase 1 & Planned & 716 & OneWeb & 51 & 1,200 \\
    \midrule
    One Web Phase 2 & Planned & 6,372 & OneWeb & 51 & 1,200 \\
    \bottomrule
    \\
    \multicolumn{6}{p{\textwidth}}{
        Note: Due to limited public information, the numbers here come from a variety of sources, or represent best estimations from available public filings. See Appendix \ref{appendix:satDetails} for more details.
    }
    \end{tabular*}
    \caption{Constellation Details}
    \label{tab:constellationDetails}
\end{table*}

\subsubsection{Inmarsat}

The victim group of GEO satellites chosen belong to the Inmarsat Global Xpress Series. These were chosen as they represent an existing communications network, exemplify well the `Old Space' approach of few, powerful satellites, and operate in the same frequency bands as Starlink and OneWeb (Ku/Ka band). 

Inmarsat do not release much information about the power of their satellites, however the limited information on the 2, 3, 4 series as well as the then upcoming Alphasat (which became Inmarsat-4A F4) give the maximum EIRP link of 70 dBW\cite{inmarsatDetails}. 


The orbits for the 5 satellites comprising the GX series were taken from Celestrak TLEs \cite{celestrak}.

\subsubsection{Constellations}

The simulation accepts both existing Two-Line Element sets (TLEs), or constellations specified via the GUI or scripting language. TLEs are a format of storing satellite positions and orbits, allowing tracking. For the paper, a number of different constellations (\cite{PlanetLabsTLEs, StarlinkAttachment, OneWebFCC}) were modelled as the interfering party, using: 

\begin{itemize}
    \item PlanetLabs' Dove Constellation  \label{item:PlanetLabsDove}
    \item CubeSat Walker Constellation \label{item:CubeSatWalker}
    \item Starlink  \label{item:StarlinkFirstGroup}
    \item Starlink Phase 1  \label{item:StarlinkPhase1}
    \item One Web Phase 1  \label{item:OneWebPhase1}
    \item One Web Phase 2 \label{item:OneWebPhase2}
\end{itemize}

Further details can be found in Table~\ref{tab:constellationDetails}.

\subsubsection{Simulation Steps}

Once the groundstations, lower constellation, and higher constellation have been specified, the simulation can be run with a number of other variable parameters. For the results below, the simulation was run with a propagation step size of 10 seconds, and a total elapsed time of 24 hours.

At every given step, each groundstation calculates all LEO satellites visible (determined by greater than 15 degrees above the horizon). It also calculates which GEO satellite has the strongest received signal. For each of these it stores distance, azimuth, and elevation for the model discussed in Section \ref{sec:Model}.

The advantage of separating the orbital simulation from the RF simulation is that a single orbital simulation can be used repeatedly by the RF model, varying parameters such as power and frequency, and changing the satellite selection process (e.g. elevation over a specific angle). This dramatically reduces the amount of repeated computation required for each total run of the pipeline. 

To further improve efficiency, when processing the RF model, we can calculate only the distance based factor of the Free Space Path Loss, and add the constant and frequency components afterwards, as see in Eq. \ref{eq:FSPL}

\begin{align}
    \label{eq:FSPL}
    FSPL &= 20 \cdot \log \left( \dfrac{4 \pi \cdot 1000 \cdot d \cdot f}{c}  \right) \\
         &= 20 \cdot \log \left( d \right) + 
            20 \cdot \log \left( \dfrac{4 \pi \cdot 1000}{c} \right) +
            20 \cdot \log \left( f \right) \\
        &= 20 \cdot \log \left( d \right) - 
           87.55221678 +
           20 \cdot \log \left( f \right)
\end{align}

where $d = \text{Distance in km}$. This allows us to calculate the `contribution' of each satellite once, based on FSPL and user terminal antenna gain, and vary the frequency and power used by the satellites afterwards to determine final interference (as used in Fig.~\ref{fig:Starlink-SINR-curves}).

Both the orbital propagation and RF simulation code will be made publicly available at \url{https://github.com/ssloxford} to assist additional security research into satellite interference in future.

\subsection{Results}

With no interference, the average received signal strength from Inmarsat was $-96.02$ dBW, with an SINR of $23.98$ dBW. However, with the added interference from each constellation considered independently, we see the results in \ref{tab:results}.

\begin{table}[ht]
\caption{Mean Interference Results}
\label{tab:results}
\setlength\tabcolsep{1.5pt}
\begin{tabular}{p{0.22\linewidth}x{0.25\linewidth}x{0.28\linewidth}x{0.2\linewidth}}
\toprule
Name              & Mean SINR (dBW) & Mean Time Jamming (\%) & Mean period of jamming (s)     \\ 
\midrule
Dove              & 23.94           &  0  & 0  \\
CubeSat           & 23.96           &  0  & 0  \\
Starlink          & 13.27           &  12.41 & 31.25  \\
Starlink Phase 1  & 9.80            &  27.35 & 38.88  \\
OneWeb            & 15.56           &  7.36  & 47.52  \\
OneWeb Phase 2    & 3.81            &  77.67 & 150.67 \\
\bottomrule
\end{tabular}
\setlength\tabcolsep{6pt}

\end{table}

However, as we will see from analysing individual ground stations, these averages obscure a large amount of information, and do not necessarily paint an accurate picture.

\subsubsection{Feasibility of Denial of Service}

\begin{figure*}[t]
    \begin{center}
        \begin{subfigure}{0.45\linewidth}
            \includegraphics[scale=0.5]{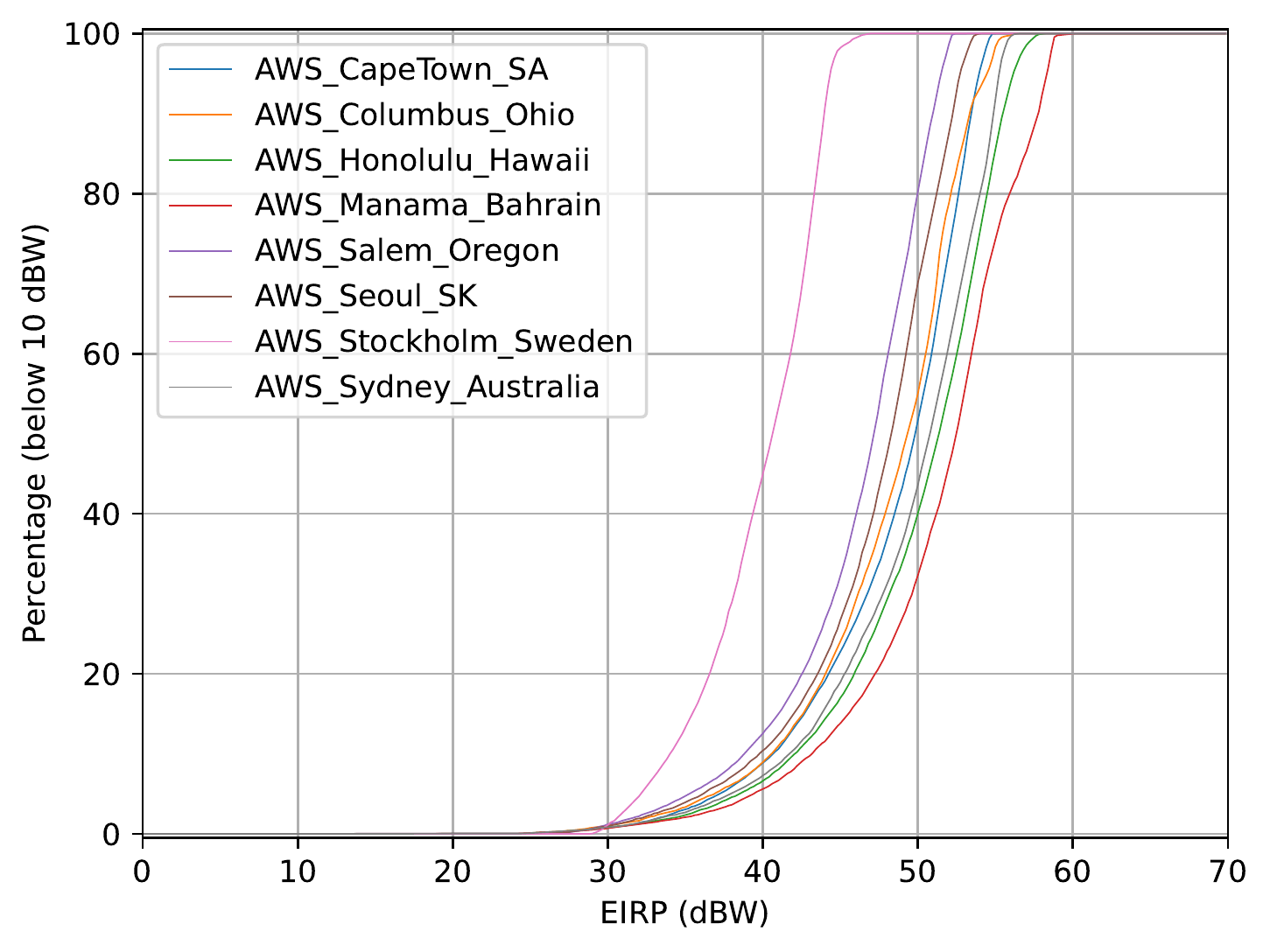}
            \caption{Starlink}
            \label{fig:Starlink-SINR-curves}
        \end{subfigure}
        \begin{subfigure}{0.45\linewidth}
            \includegraphics[scale=0.5]{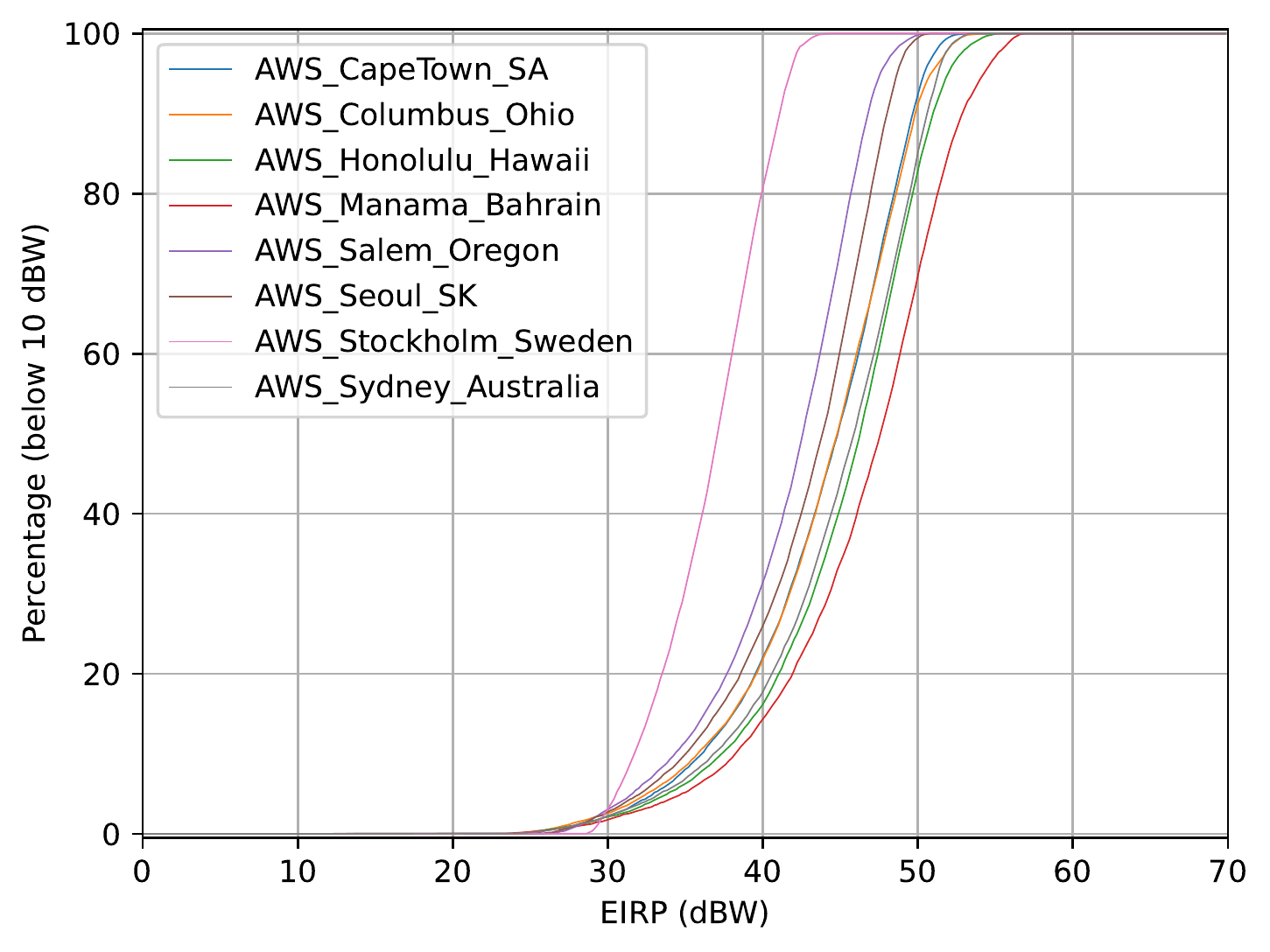}
            \caption{Starlink Phase 1}
            \label{fig:StarlinkPhase1-SINR-curves}
        \end{subfigure}
    \end{center}
    \caption{Time spent jamming over 24 hours as a function of the power of each satellite in the constellation. Starlink Phase 1 has triple the number of satellites, which reduces the per-satellite power required to achieve the same jamming.}
\end{figure*}

Here `jamming' is defined as bringing the SINR below 10 dBW. This specific number is somewhat arbitrary\footnote{Most services do not publish such low-level numbers, but one service lists 10 dBW as required to show correct alignment \cite{BigBlu}.}, but allows us to set some threshold for analysis.

Fig.~\ref{fig:Starlink-SINR-curves} shows us how often the attacker could expect to be jamming as the power of the satellite scales. For Starlink, with EIRP of 39 dBW, it means that for all groundstations barring Stockholm there would be disruption $5-15\%$ of the time. Jamming is clearly feasible then, and affected greatly by the individual power of each satellite though as constellations of greater number are considered, such as Starlink Phase 1, the number of satellites can make up for lack of individual power (see Fig.~\ref{fig:StarlinkPhase1-SINR-curves}) on the order of roughly 5 dBW (or roughly 3.16 times less power).

\subsubsection{Effect of Groundstation Location}

From the groundstation curves we can see that the non-uniform layout of the constellation gives different effectiveness at different groundstations. Even if the curves between stations are similar, their slight offset and rapid growth means that for the same power level they might experience very different jamming results.

\begin{figure}[t]
\begin{center}
\includegraphics[scale=0.55]{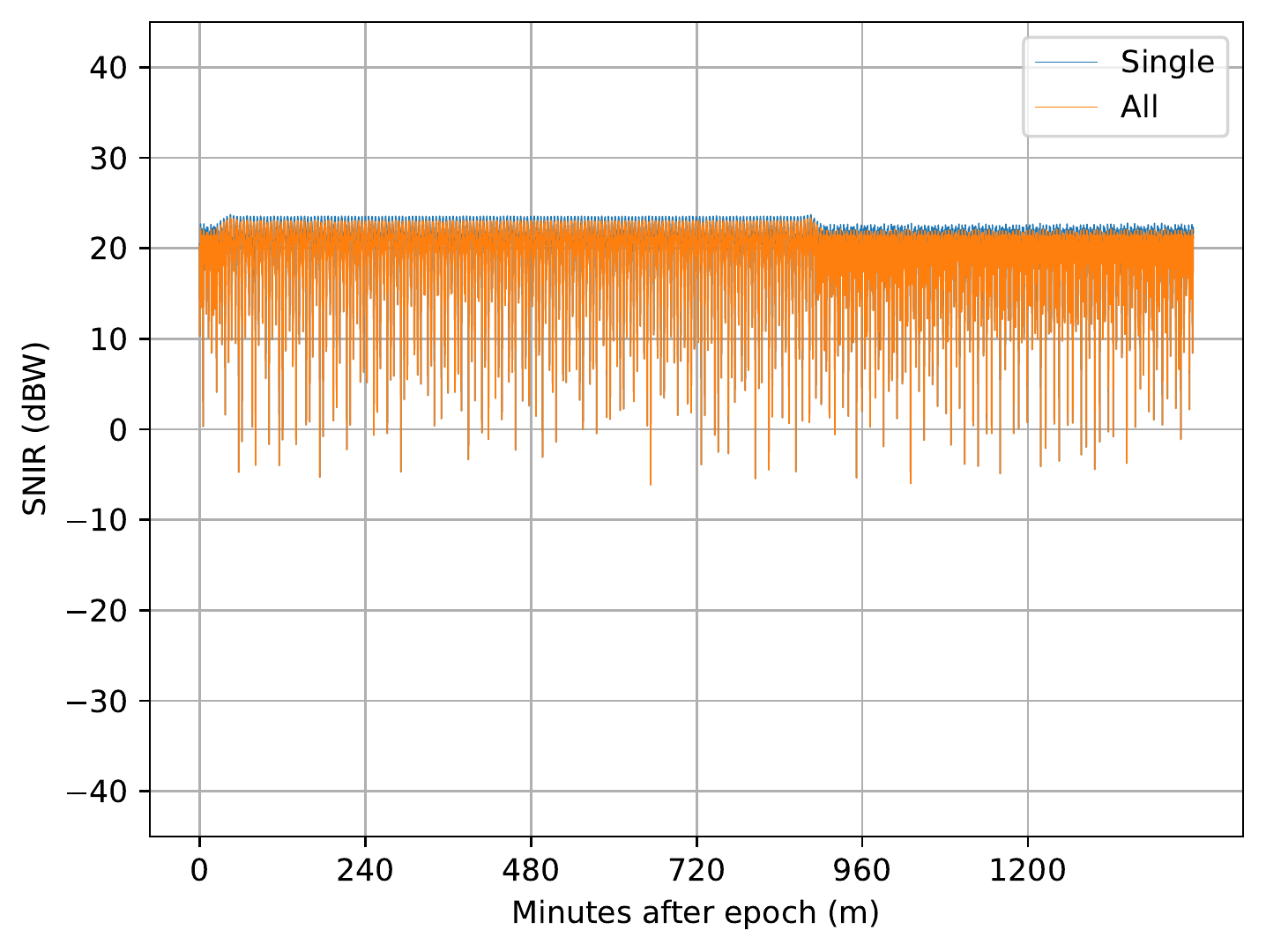}
\caption{Starlink SINR at Manama over 24 hours. Every point represents 10 seconds, and each spike below 10 dBW is considered to be jamming.}
\label{fig:Starlink-AWS_Manama_Bahrain-SINR}
\end{center}
\end{figure}

\begin{figure}[t]
    \begin{center}
    \includegraphics[scale=0.55]{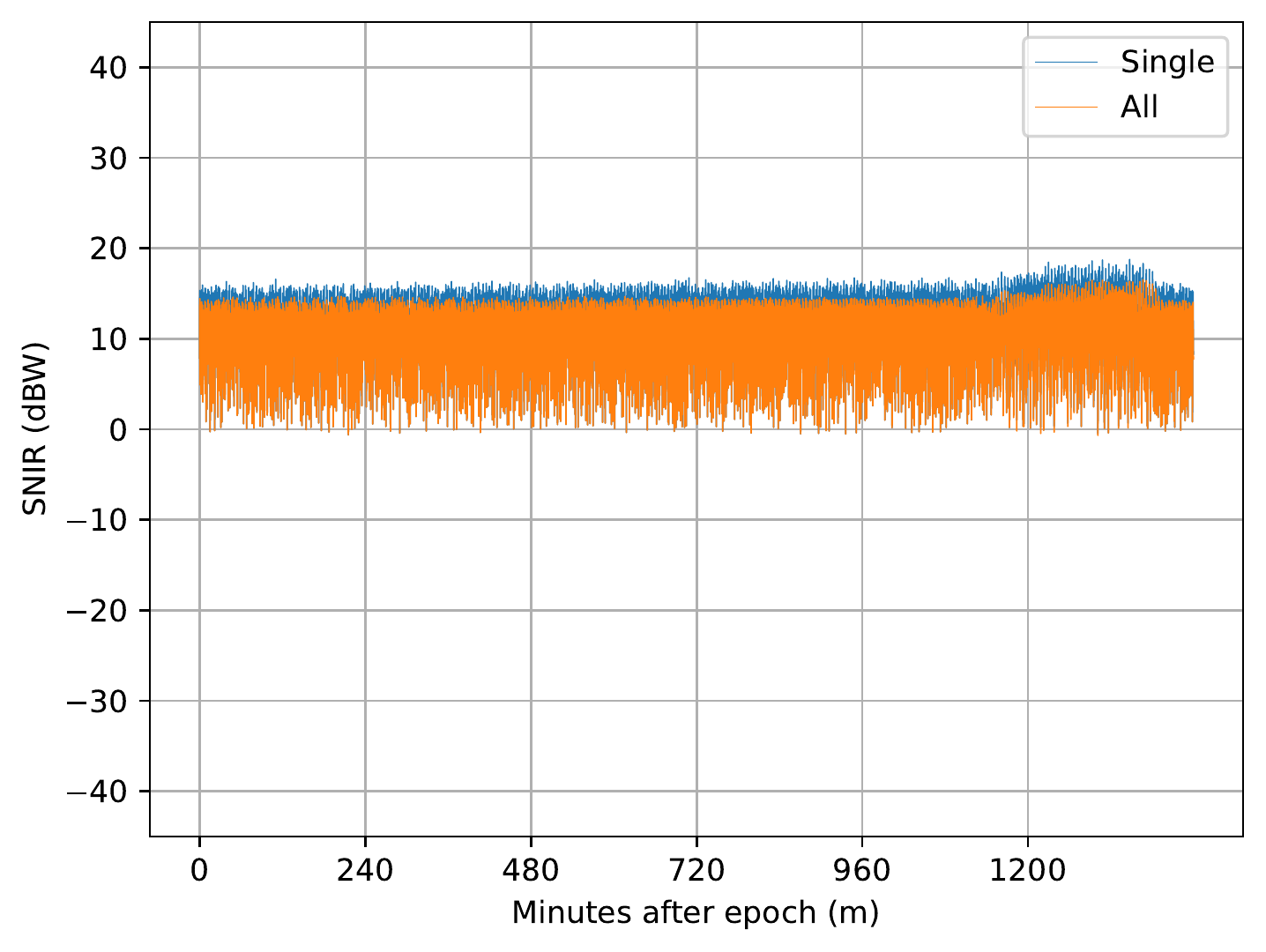}
    \caption{Starlink SINR at Stockholm over 24 hours}
    \label{fig:Starlink-AWS_Stockholm_Sweden-SINR}
\end{center}
\end{figure}

For example, comparing the SINR directly over the course of the 24 hours between Manama and Stockholm we can see the sporadic coverage given to Manama in Fig.~\ref{fig:Starlink-AWS_Manama_Bahrain-SINR} compared to the denser coverage over Stockholm (Fig.~\ref{fig:Starlink-AWS_Stockholm_Sweden-SINR}), an effect noted in \cite{Cofrequency}. Due to its inclination of $53.2\si{\degree}$, Starlink has a much higher density near groundstations like Dublin and Stockholm ($53.3\si{\degree}$ and $59.3\si{\degree}$ latitude respectively), than the likes of Honolulu and Manama ($21.3\si{\degree}$ and $26.2\si{\degree}$ latitude respectively). This shows how important specific orbital parameters, as well as the locations an attacker wishes to jam, are in determining the potential jamming effectiveness of a constellation. A satellite's position relative to the vector between groundstation and victim target (and how often it intercepts this vector) can matter much more than the satellite's power.

However satellite power is not to be discounted. There is a clear minimum power required on all curves seen, and after that it looks like linear order growth (though shown on a logarithmic scale), until it plateaus at the very end. Doubling in power therefore approximately doubles the time spent jamming.

\subsubsection{Effect of Constellation Choice}

However, not all constellations exhibit the patterns shown above. OneWeb, for example, has a much more cyclic and predictable nature.

\begin{figure}[t]
\begin{center}
\includegraphics[scale=0.55]{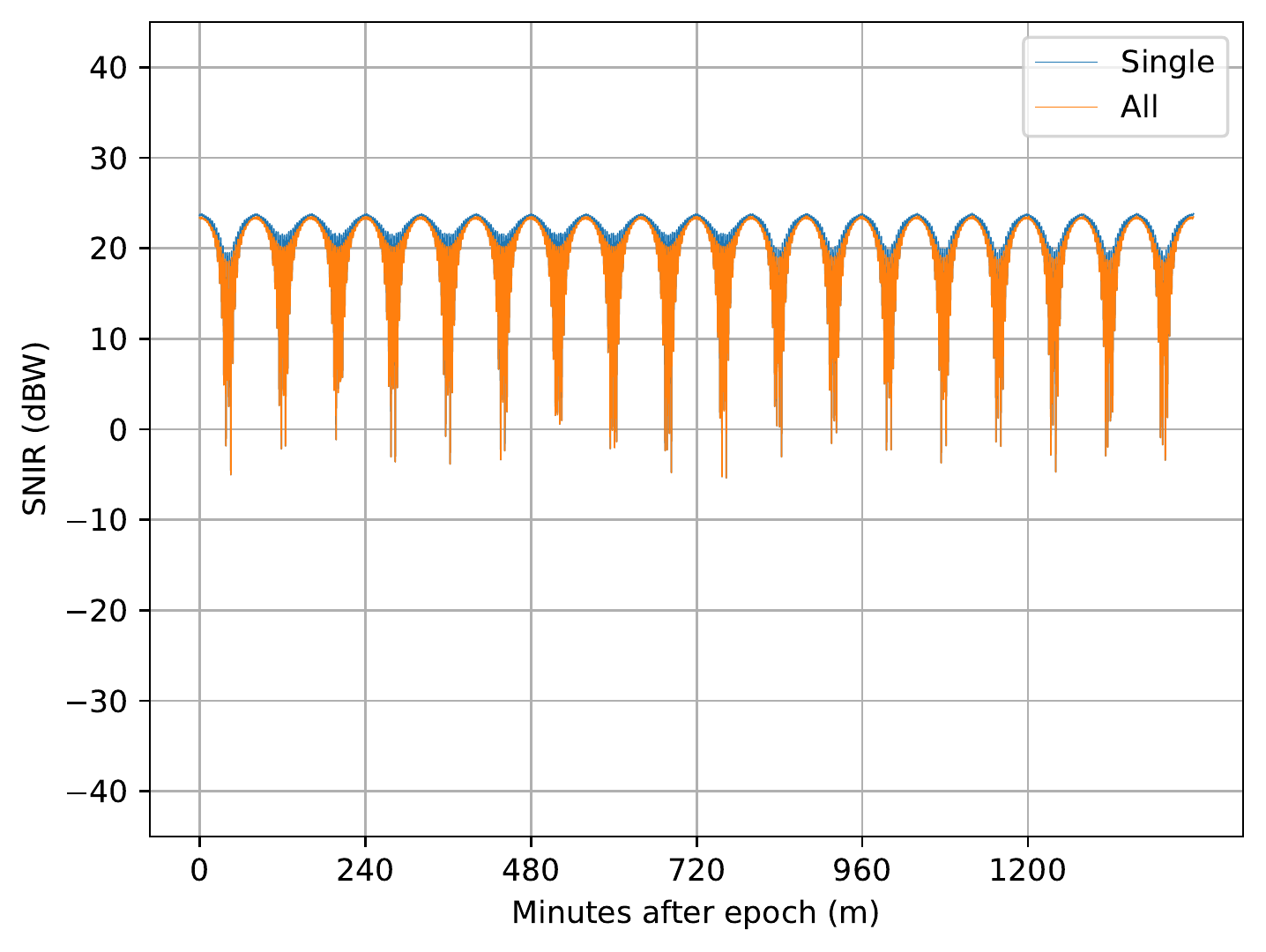}
\caption{OneWeb SINR at Manama over 24 hours}
\label{fig:OneWeb-AWS_Manama_Bahrain-SINR}
\end{center}
\end{figure}

We can see that less time is spent jamming overall, but it is much clearer when that happens, and easier to predict. 

\begin{figure}[t]
    \begin{center}
    \includegraphics[scale=0.55]{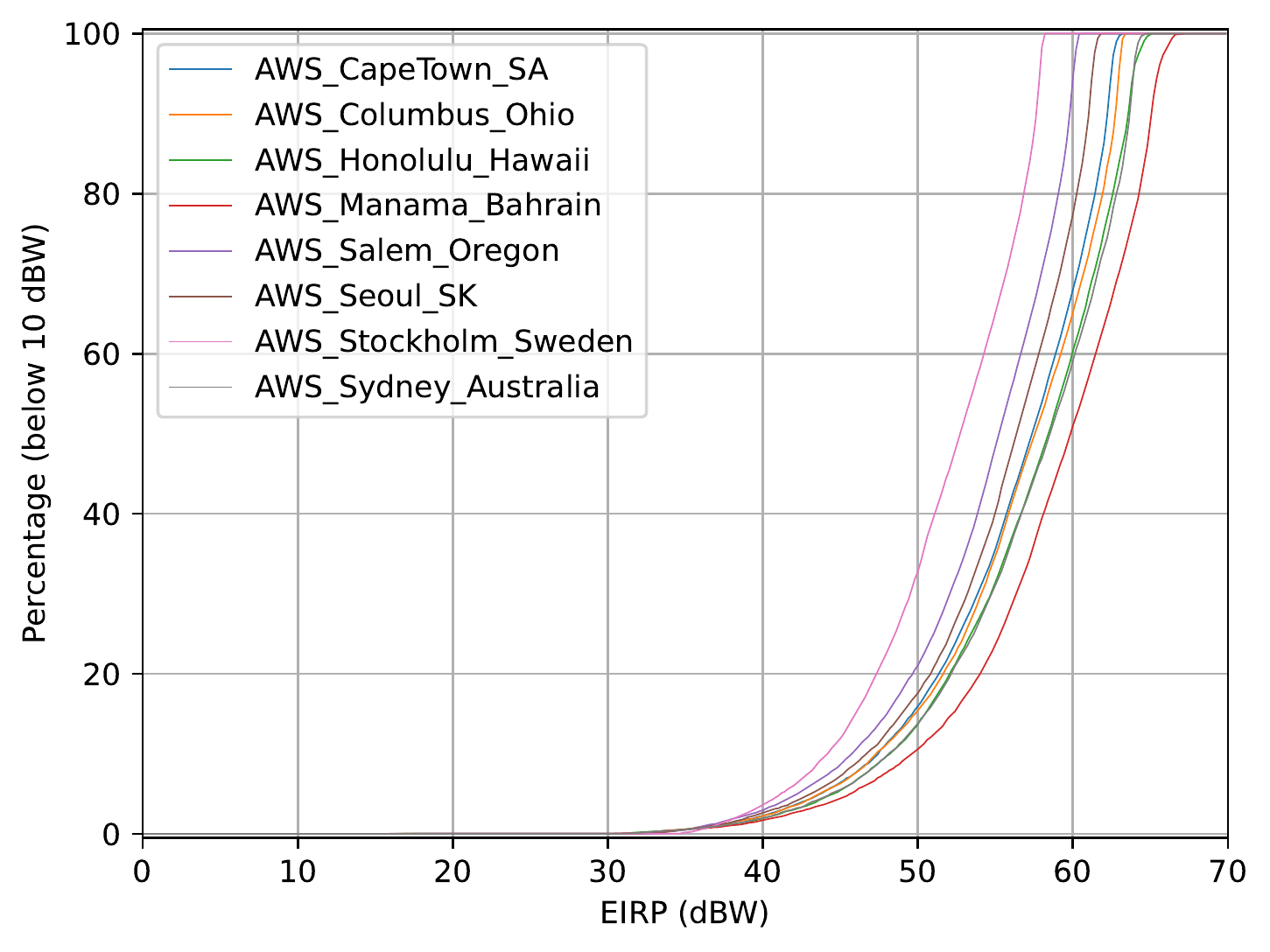}
    \caption[]{Groundstation curves for OneWeb with a threshold of 10 dBW}
    \label{fig:OneWeb-SNIR-curves-10}
    \end{center}
\end{figure}

Looking at time jamming across all ground stations for OneWeb as seen in Fig.~\ref{fig:OneWeb-SNIR-curves-10} the lower number of satellites, higher orbit, and strict periodic nature of the jamming leaves it much less effective than Starlink, despite the higher power of each satellite.

\begin{figure}[t]
    \begin{center}
    \includegraphics[scale=0.55]{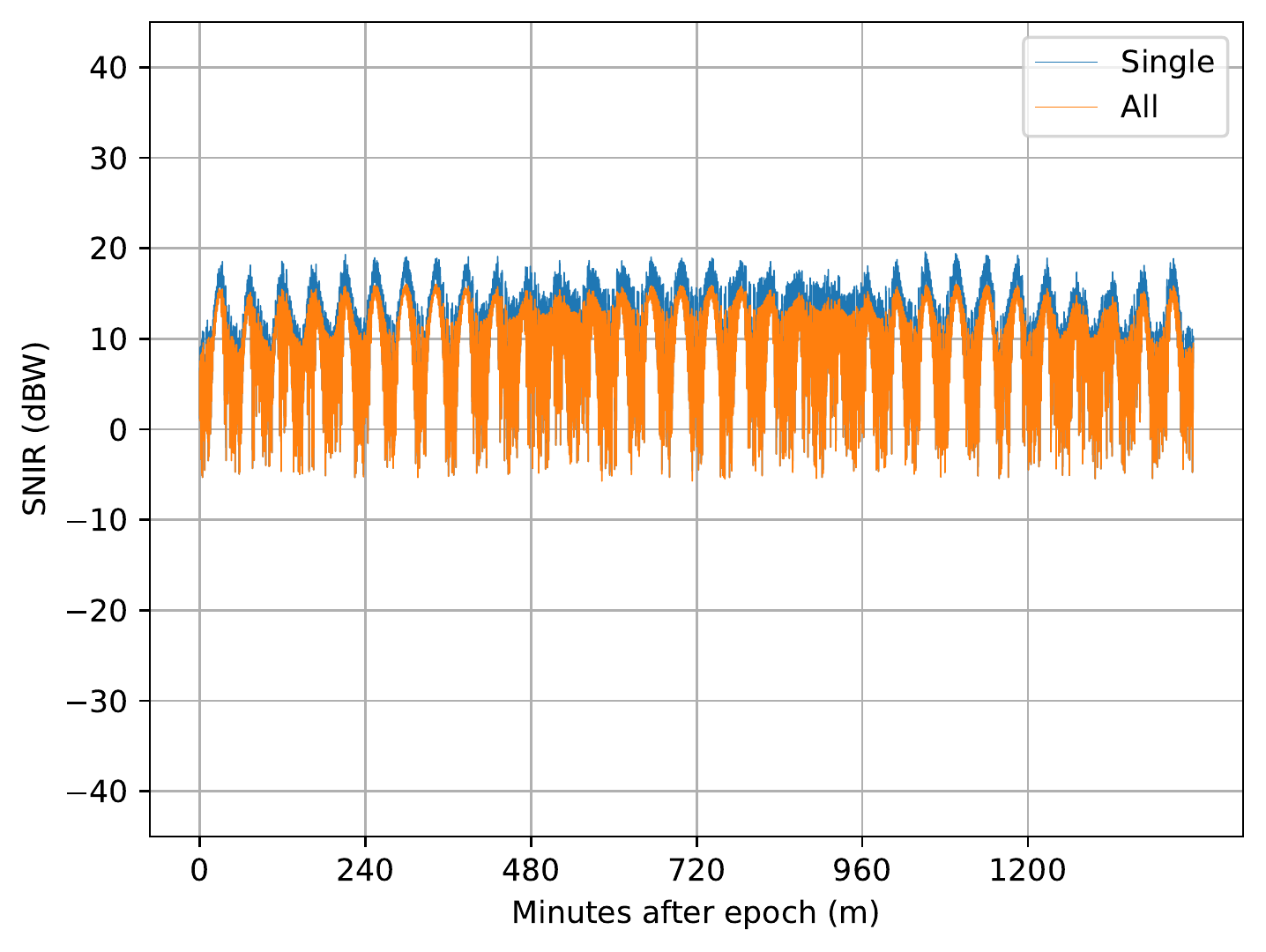}
    \caption{OneWebPhase2 SINR at Manama over 24 hours}
    \label{fig:OneWebPhase2-AWS_Manama_Bahrain-SINR}
    \end{center}
\end{figure}

Contrasting constellations based on density, such as between OneWeb (Fig.~\ref{fig:OneWeb-AWS_Manama_Bahrain-SINR}) and OneWeb Phase 2 (Fig.~\ref{fig:OneWebPhase2-AWS_Manama_Bahrain-SINR}), we can also see how much of a difference the `secondary' satellites make (those not directly between the groundstation and victim satellite). The greatest difference between the single strongest interfering satellite, and the sum of all interfering satellites is $\approx 5$ dBW for One Web Phase 2.

\subsubsection{CubeSat vs. Commercial New-Space}

So far we have still been considering constellations with hundreds or thousands of relatively powerful satellites however. Looking at the other end of the spectrum, at a much more achievable CubeSat constellation, the results are very different.

\begin{figure}[t]
    \begin{center}
    \includegraphics[scale=0.55]{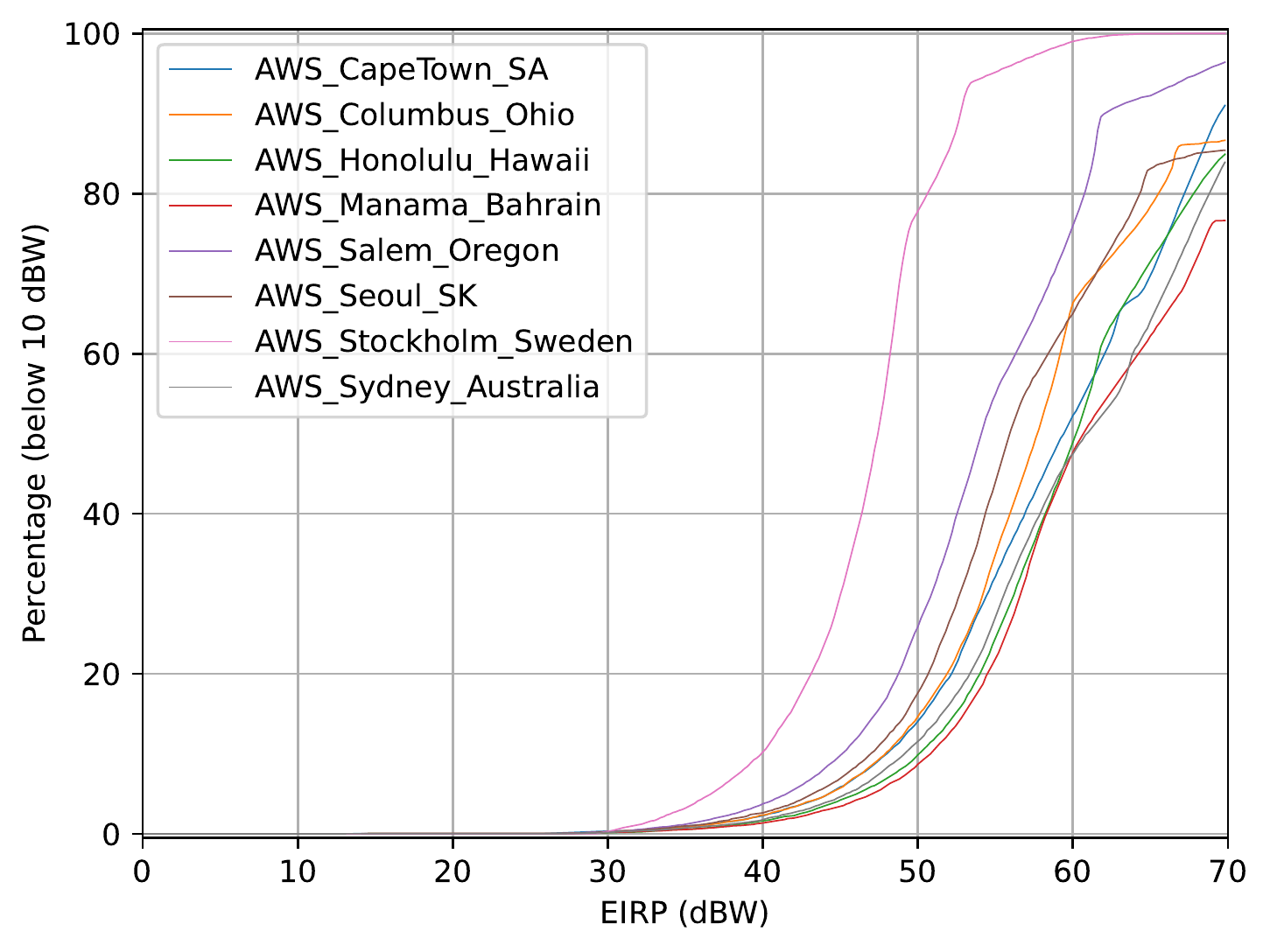}
    \caption[]{Groundstation curves for a CubeSat Walker Constellation with a threshold of 10 dBW}
    \label{fig:walker11x36-SNIR-curves-10}
    \end{center}
\end{figure}

The CubeSat constellation (Fig.~\ref{fig:walker11x36-SNIR-curves-10}) itself has better performance compared to OneWeb (Fig.~\ref{fig:OneWeb-SNIR-curves-10}), with the notable improvement of Stockholm (again explained by the lower inclination), until a power of 60 dBW. This is currently far beyond the power achievable at such a scale, and so matters much less in comparison. The crucial difference however the actual EIRP of each CubeSat sits at 6 dBW, where no groundstation is being jammed at all. Taking an example groundstation, in this case Salem, we can see that the constellation never even comes close to appreciably reducing the SINR.

\begin{figure}[t]
\begin{center}
\includegraphics[scale=0.55]{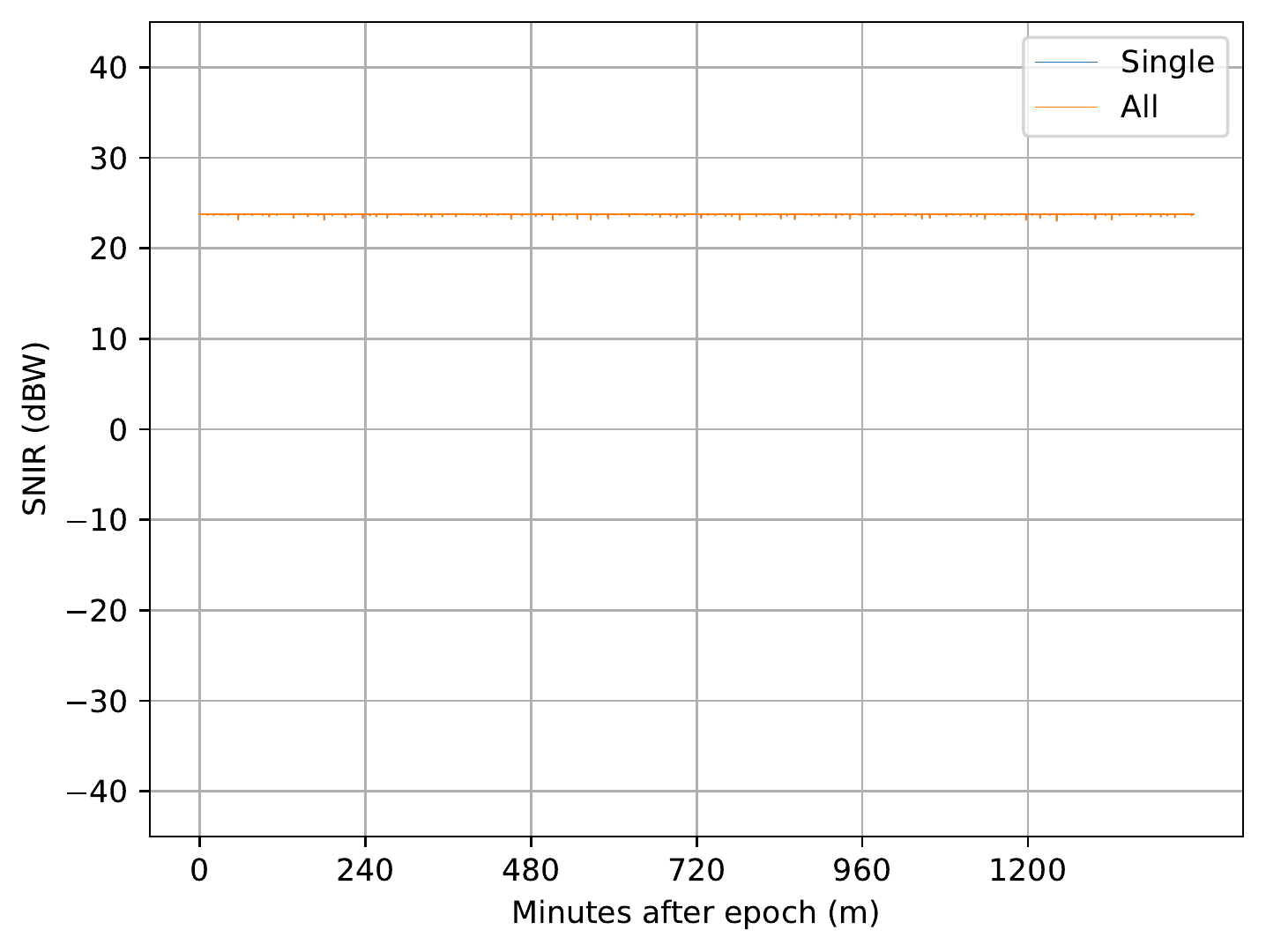}
\caption{CubeSat SINR at Salem over 24 hours}
\label{fig:walker11x36-AWS_Salem_Oregon-SINR}
\end{center}
\end{figure}

\section{Discussion}
\label{sec:Discussion}

The results give a number of interesting considerations. For one, the latitude makes a large difference, with groundstations closer to the equator suffering less interference than those higher up. This is chiefly because constellations with a non-polar inclination (such as Starlink) are much denser at the latitude of their inclination (53.2 degrees). This means that there are more secondary satellites contributing, and the primary satellites spend more time with low angular separation. 

Next we can see that the nonlinearity of the SNIR curves (such as Fig.~\ref{fig:Starlink-SINR-curves}) means a relatively small change in power can result in a dramatic increase in jamming time. Satellite manufacturers are often very restrictive in the information they provide about their satellites. As such, lots of the information used is taken from legal filings. This has the unfortunate consequence that all of the information provided is what the manufacturers \textit{intend} to do, not their full capabilities (which is what we care about in an adversarial scenario). There are numerous references to lowering satellites' EIRP to acceptable levels, which implies that they are more capable than the given figures suggest \cite{StarlinkAttachment}. This makes sense, given that the satellite operators are continually trying to seek relaxation of regulation. If the satellites are launched with a higher maximum EIRP, they can adopt it as soon as legally permitted, rather than building and launching satellites with the updated specifications. As such, jamming is likely to be more possible than considered so far, which represents a conservative estimate.

\begin{figure}[h]
    \begin{center}
    \includegraphics[scale=0.55]{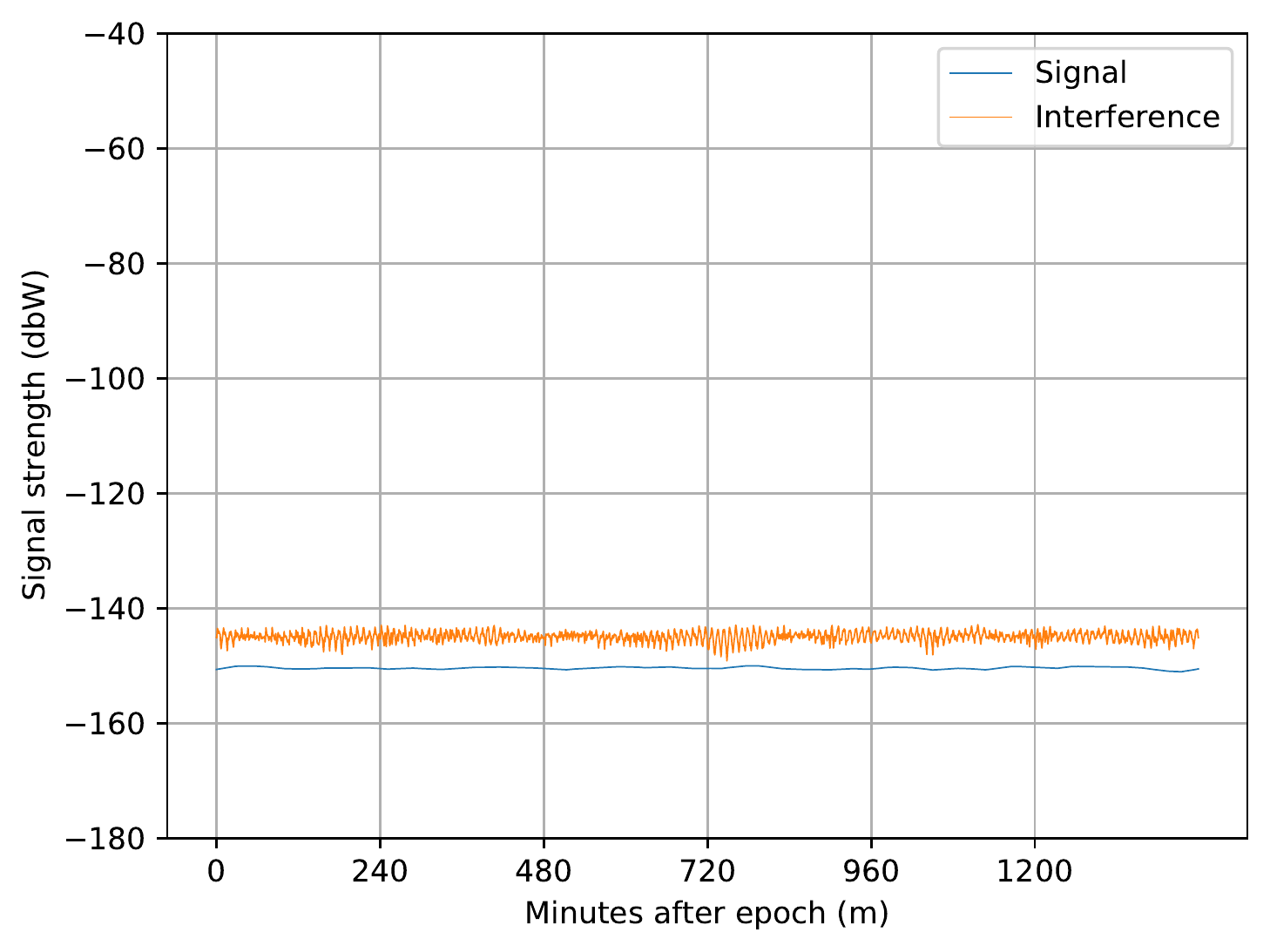}
    \caption[]{A comparison between average simulated GPS signal strength (blue) compared to average CubeSat signal strength.}
    \label{fig:walker11x36-GNSS}
    \end{center}
\end{figure}

Finally, the results show that CubeSats simply do not have the power required to compete with most Ka-band services. However, communications are not the only space-based service in use, Global Navigation Satellite Systems (GNSS) being a key alternative. The European Space Agency (ESA) and others use S-Band (2220-2290 MHz down) for Telemetry, Tracking, and Command (TT\&C) signals. This is relatively close to the L1-Band (1575.42 MHz) used by GNSS. A COTS SDR (for example the TOTEM Nanosatellite SDR Platform \cite{TOTEM}) covers 70-6000 MHz, and therefore a satellite could reasonably use this SDR for operational or potentially malicious purposes. While interference wouldn't be effective, since GNSS is built to deal with extracting a weak signal from background noise, it does still leave open the option of a replay or spoofing attack to be explored in future work, since when simulated the CubeSat constellation still gives a more powerful signal than GNSS as seen in Fig.~\ref{fig:walker11x36-GNSS}.

\begin{figure*}[t]
    \begin{center}
        \subfloat[Starlink vs Starlink Phase 1]{\includegraphics[scale=0.55]{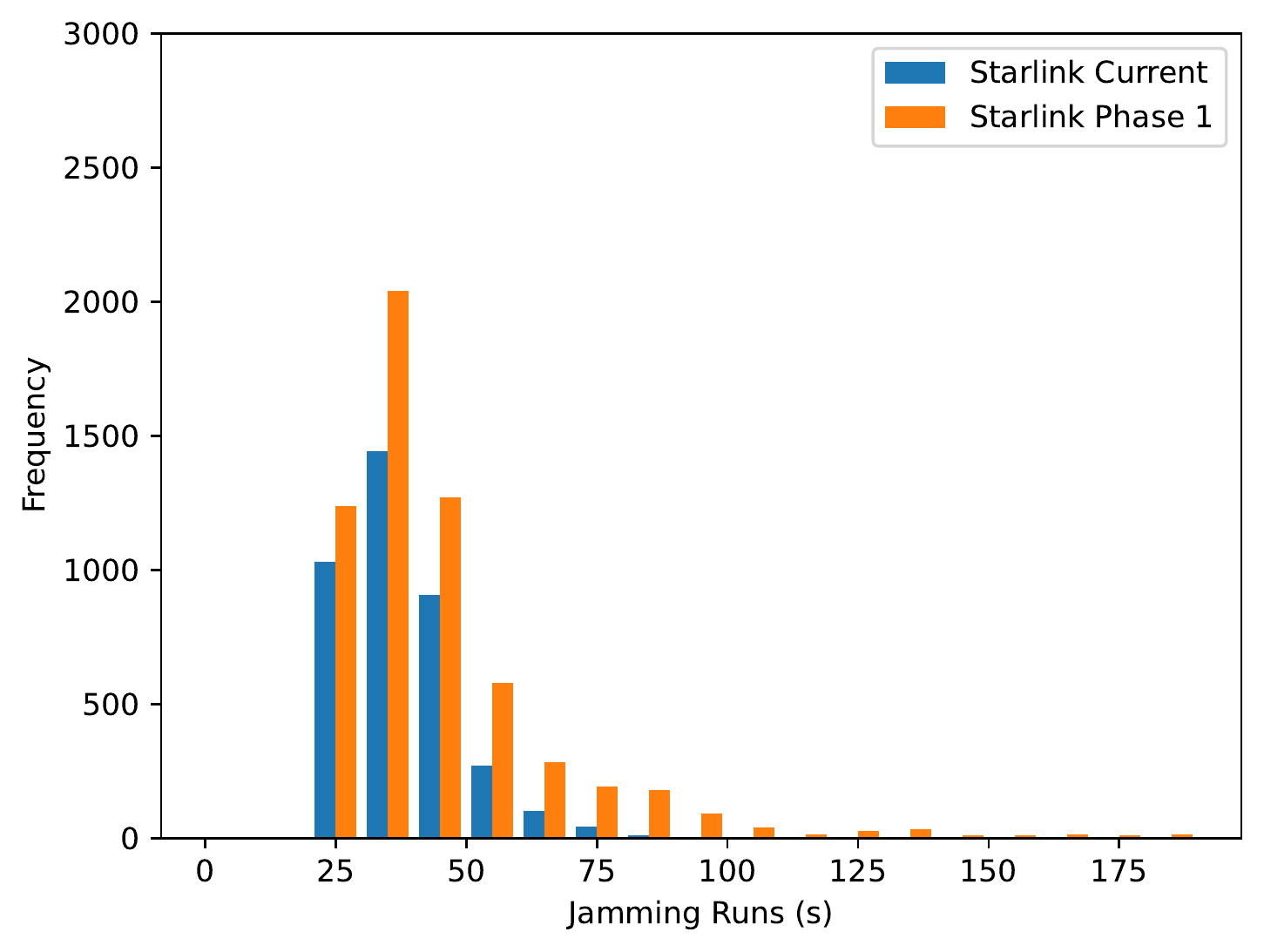}
        \label{fig:Starlink-Hist}}
        \subfloat[OneWeb vs OneWeb Phase 2]{\includegraphics[scale=0.55]{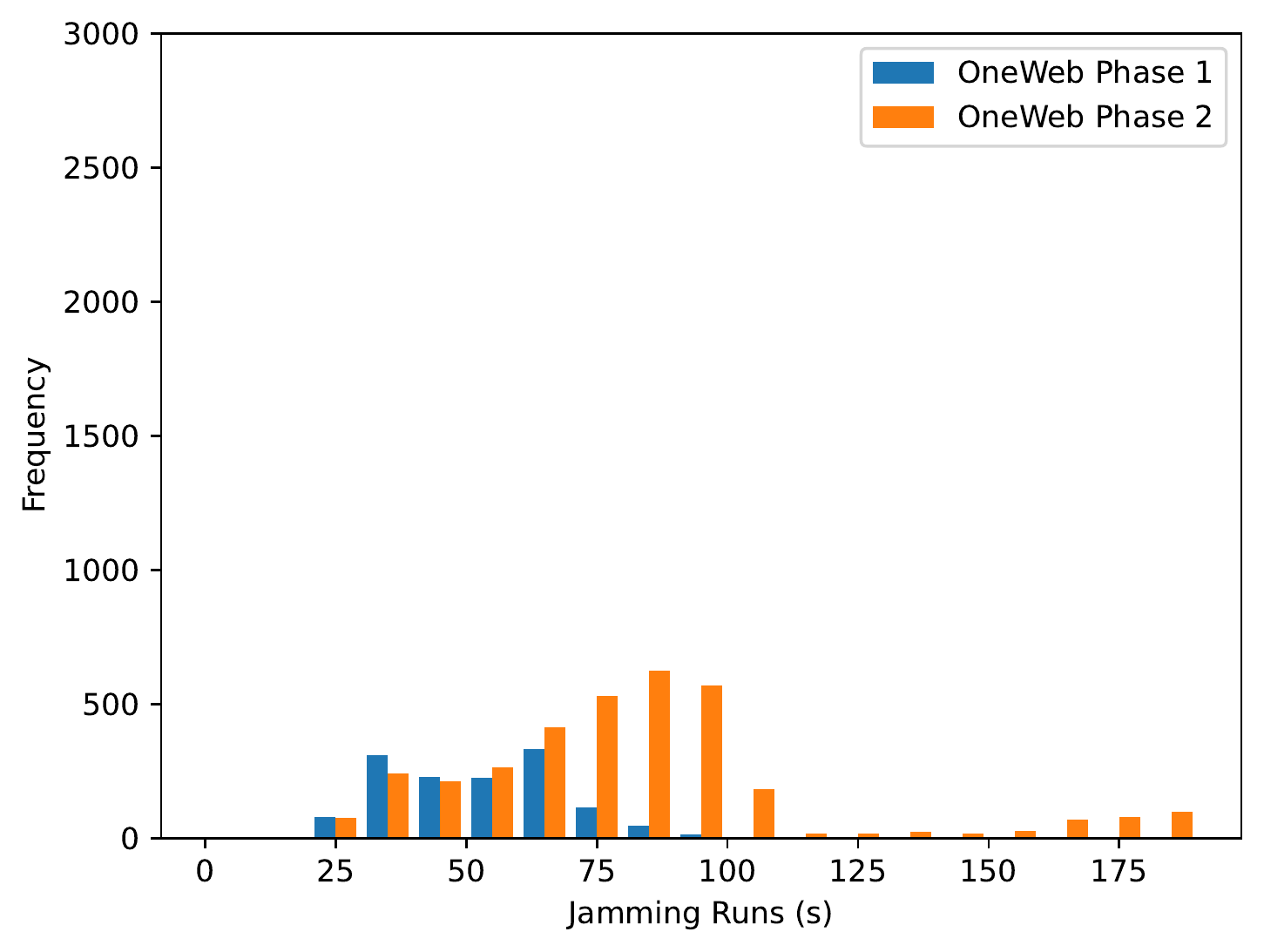}
        \label{fig:OneWeb-Hist}} 
    \caption{Figures showing histograms of `jamming runs'. These show the frequencies of time periods of constant interference (SNIR $< 10$ dBW).}
    \label{fig:Constellation-Hist}
    \end{center}
\end{figure*}

\subsection{Usage Analysis}
\label{sec:UsageAnalysis}

We have seen that constellation design plays a large part in the type of jamming available, with Starlink and the CubeSats having frequent but seemingly random spikes, contrasted with OneWeb and more PlanetLabs' periodic jamming. This could well be an advantage for an attacker, knowing when a given system might be down (if only briefly). However if the user/defender is attentive to this pattern, and does not require 100\% uptime, then it may be very easy for the defender to adapt and instead use the system outside of these times. One disadvantage for the attacker is that they are completely constrained by orbital dynamics, and so cannot alter when they jam (though can choose to jam less than maximally possible). This will clearly work best if the defender can't choose to use the system outside of the times of jamming.

We can look more closely at how long each constellation is able to jam for by plotting a histogram of jamming time across all groundstations. Figure \ref{fig:Constellation-Hist} shows this distribution for both Starlink constellations, as well as both OneWeb constellations. We can see in Fig~\ref{fig:Starlink-Hist}, Starlink never jams for longer than 90 seconds. This means that as long as the timeout period for packets is longer than 90 seconds, everything should eventually get through. This works for downloading files, but would render real-time communication via video or speech unusable. Uploads could in theory remain unaffected, however the TCP protocol requires acknowledgment within a certain window\cite{TCPextension}, and if the interference continues at sufficient strength it might not receive this. In that case, even uploading would be stopped until connection resumes.

However, Starlink Phase 1 has (limited) interference capability up to almost 3 minutes, and OneWeb Phase 2 has an average period of 2 and half minutes, happening frequently. This presents a very serious problem for almost any use-case. Remotely managing systems with consistent two minute delays dramatically lengthens control feedback loops (i.e. between triggering a command and seeing the effect), which in industry may have expensive or dangerous consequences, especially if the satellite communications are used for warning or monitoring systems \cite{ESAoil}.

\subsection{Attack Mitigations}
\label{sec:Mitigations}

There are a number of measures that a user on the ground might be able to take to reduce the effectiveness of interference. Chief among them are the use of a more highly directional antenna and Frequency Hopping Spread Spectrum (FHSS).

\begin{figure}[ht]
    \begin{center}
    \includegraphics[scale=0.55]{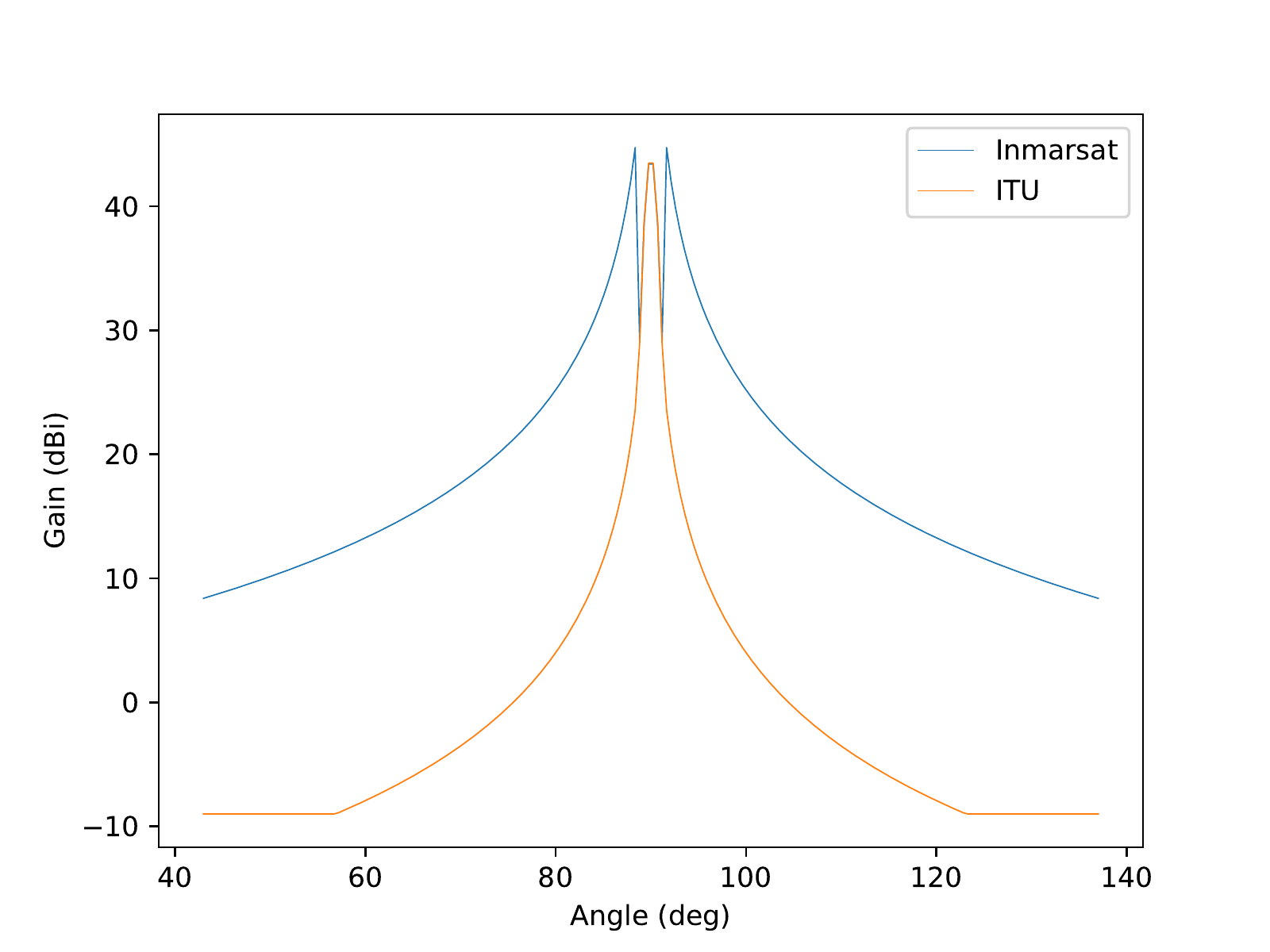}
    \caption[]{The current Inmarsat antenna gain pattern (blue), compared to an alternative recommended by the ITU (orange) that would provide much lower Interference.}
    \label{fig:ITU-Antenna-Pattern}
    \end{center}
\end{figure}

The use of more focused antenna pattern dramatically reduces not only the effect of `secondary' interfering satellites, but also minimises the amount of time a primary constellation satellite spends at a `high-gain` angle (i.e. sufficiently close to the peak gain of the antenna). This does not completely eliminate the possibility of interference, but limits how often it could cause serious disruption. An antenna pattern like that recommended by the ITU (Fig.~\ref{fig:ITU-Antenna-Pattern}) is a good starting point \cite{antennaPattern}.

Frequency hopping, the process of suddenly changing operating frequency to avoid interference, requires much deeper integration by the satellite operator than simply changing antenna. Both the transmitter (GEO satellite) and the receiver (user terminal) must be ready to change at the same time, and know which frequency they are going to next. This is extremely effective at mitigating interference, since the jamming signal from the constellation would remain at the previous frequency. This interference will therefore have a significantly reduced impact if the two frequencies are sufficiently far apart \cite{FrequencyHopping}. Frequency hopping is already commonly implemented on military systems to reduce interference, as well as in the Bluetooth protocol \cite{NATOFrequencyHopping, bluetooth}. This may be more difficult in satellite communications given the strict limits on licensed bands, but there are a number of unregulated bands that could be used for this purpose if needed. 

\section{Conclusion}


In this paper we have modeled interference from a variety of current and planned constellations at a number of groundstations. We found that jamming would already be possible, and its potential effectiveness is only going to increase as constellations grow larger. However, even the most effective constellation design considered, OneWeb Phase 2, would require a minimum EIRP per satellite of roughly 30 dBW; two orders of magnitude more powerful than the biggest CubeSats (10 dBW). For Ku/Ka band then, CubeSat interference does not seem to pose a problem, though it might have serious consequences for GNSS infrastructure. 

This power requirement means that the attack is not open to all `New Space' players, and puts it beyond the reach of the likes of all but large companies (e.g. SpaceX, OneWeb, Amazon) willing to invest the money into producing a constellation of such satellites. It is unlikely then that a constellation dedicated to this attack would be feasible. However targeting an existing or planned constellation continues to be a viable option, and the security policies of satellite operators should reflect this.

It is not enough to provide reports on avoiding interference through modulating power, direction, and operation conditions, as is currently required, as these all rely on a benign operator. To most effectively minimise their impact on other space assets, comprehensive external code reviews, adversarial analysis, and continual monitoring would all help to ensure system integrity.

The use of off-the-shelf software like Linux is excellent for rapid development and use of existing libraries, but it also lowers the barrier to entry for malicious attacks, which we have shown here can have consequences far beyond the company compromised. 

The potential mitigations mentioned in Section \ref{sec:Mitigations} go some way towards alleviating the potency of such an attack, though these rely on more capable and therefore expensive receivers, which may limit their implementation.










\appendix

\section{Appendix: Satellite Details}
\label{appendix:satDetails}

Details on satellite characteristics:

\subsection{Dove Satellite}

Planet Labs have been extremely open about both the optics and communication details of their satellites. Given the vast quantity of data produced by the cameras (80 GB per pass), EIRP is more important for these imaging satellites than might be initially expected. 

Their paper gives clear numbers for EIRP: 8.2 dBW \cite{DoveTechnical}.

\subsection{CubeSat}

For the example cubesat, we used the EXA Kratos 1U \footnote{\url{https://www.cubesatshop.com/product/kratos-1u-cubesat-platform-1-step-solution/}} Cubesat Platform, with a 6 dBW patch antenna, and 1 W transmission power, providing an EIRP of 6 dBW.

\subsection{Starlink}

Based on the FCC Starlink Attachment, the highest usable EIRP density from Starlink satellites is 15.70 dBW/4kHz  \cite{StarlinkAttachment}. Since we are assuming bandwidth of 250 MHz, that gives an EIRP $\approx 39.68$ dBW.

\subsection{OneWeb}

The OneWeb FCC filing \cite{OneWebTechnical} gives a maximum downlink EIRP density of -2.7 dBW/4kHz. Using the same bandwidth as above, that gives a maximum EIRP $\approx 45.26$ dBW.

\section{Plots for all Architectures}

\begin{figure*}[ht]
    \begin{center}
        \begin{subfigure}{.5\textwidth}
            \includegraphics[scale=0.55]{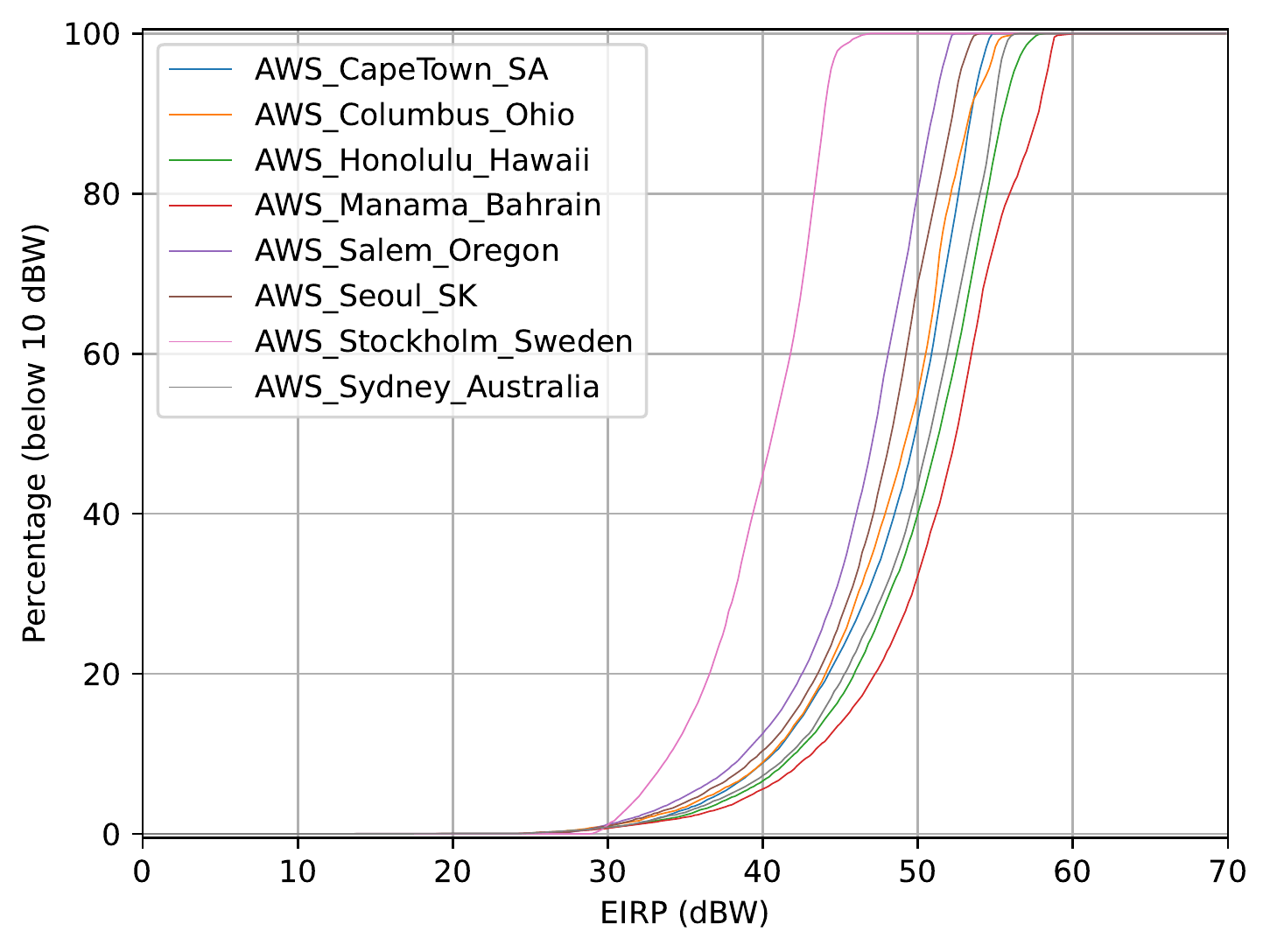}
            \caption[]{Starlink}
        \end{subfigure}%
        \begin{subfigure}{.5\textwidth}
            \includegraphics[scale=0.55]{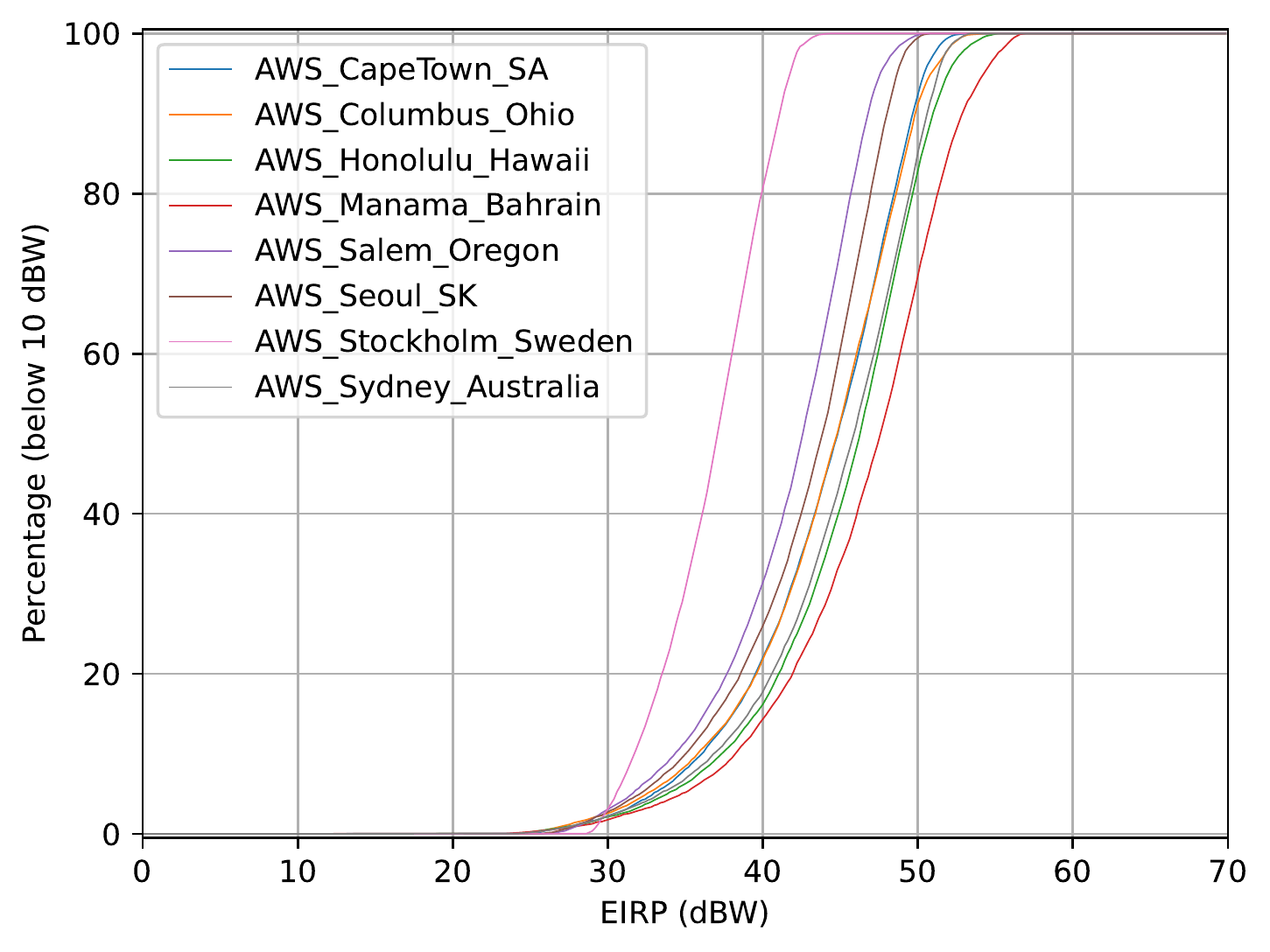}
            \caption[]{Starlink Phase 1}
        \end{subfigure}
        \begin{subfigure}{.5\textwidth}
            \includegraphics[scale=0.55]{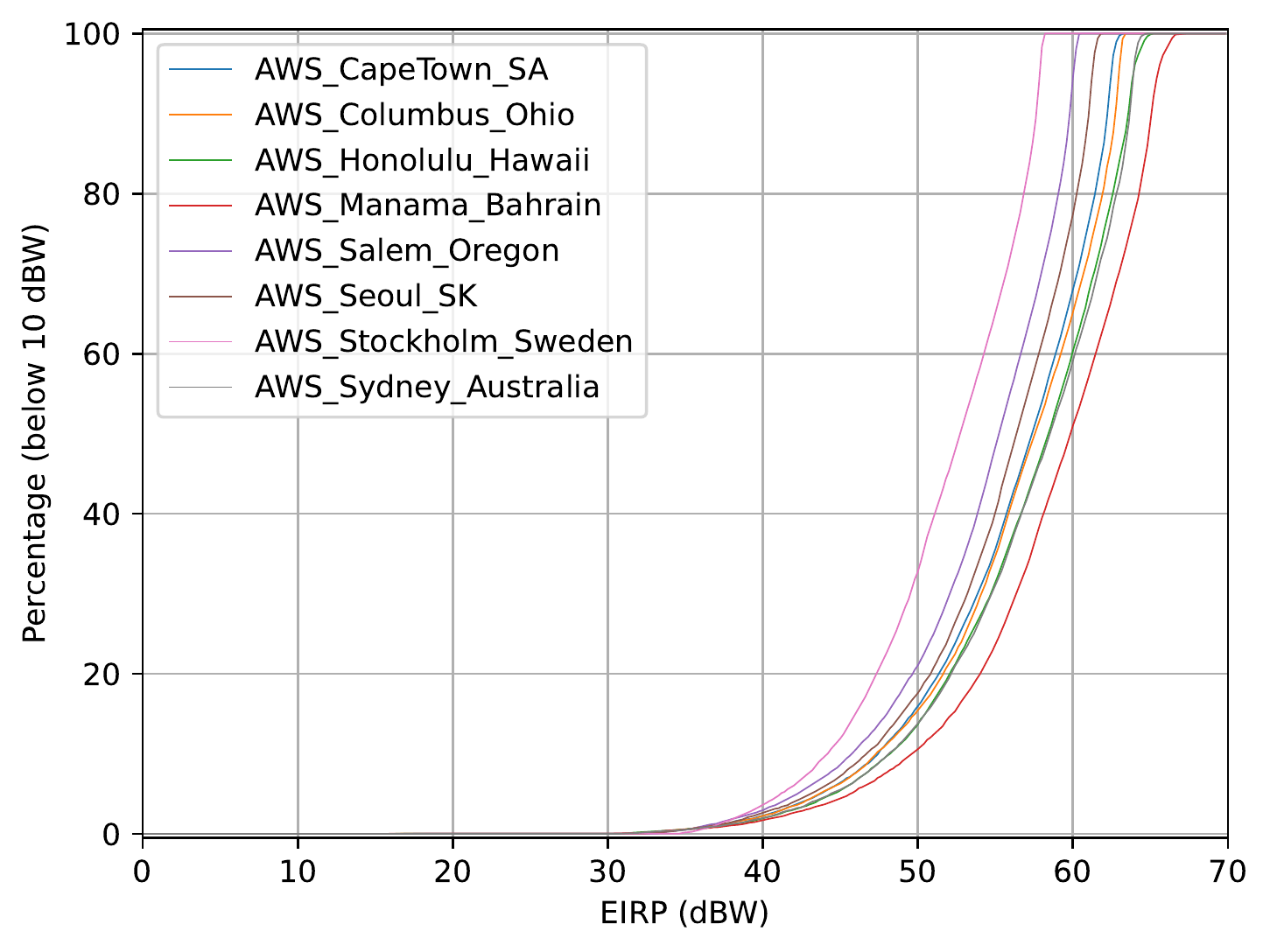}
            \caption[]{OneWeb}  
        \end{subfigure}%
        \begin{subfigure}{.5\textwidth}
            \includegraphics[scale=0.55]{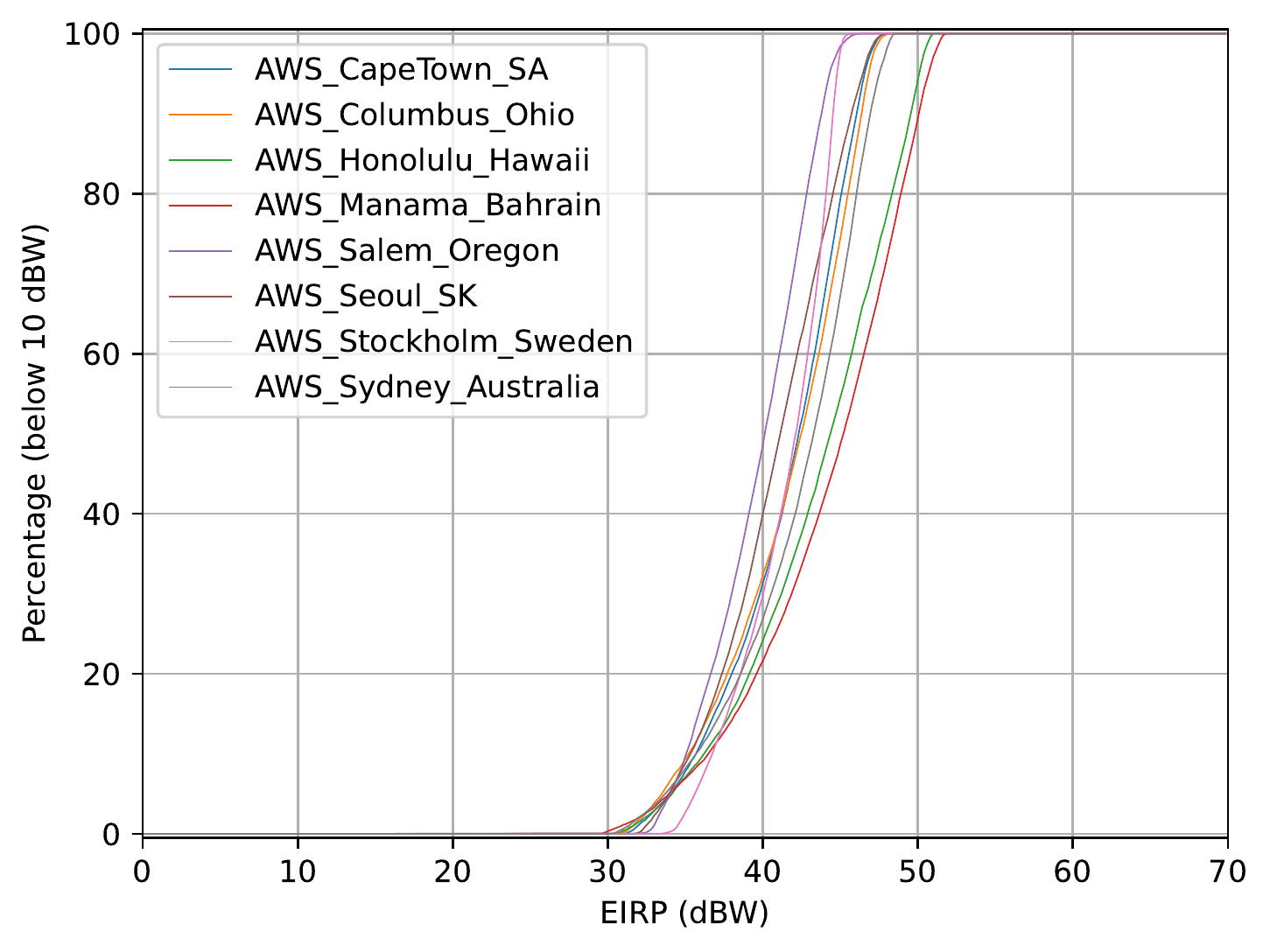}
            \caption[]{OneWeb Phase 2}
        \end{subfigure}
        \begin{subfigure}{.5\textwidth}
            \includegraphics[scale=0.55]{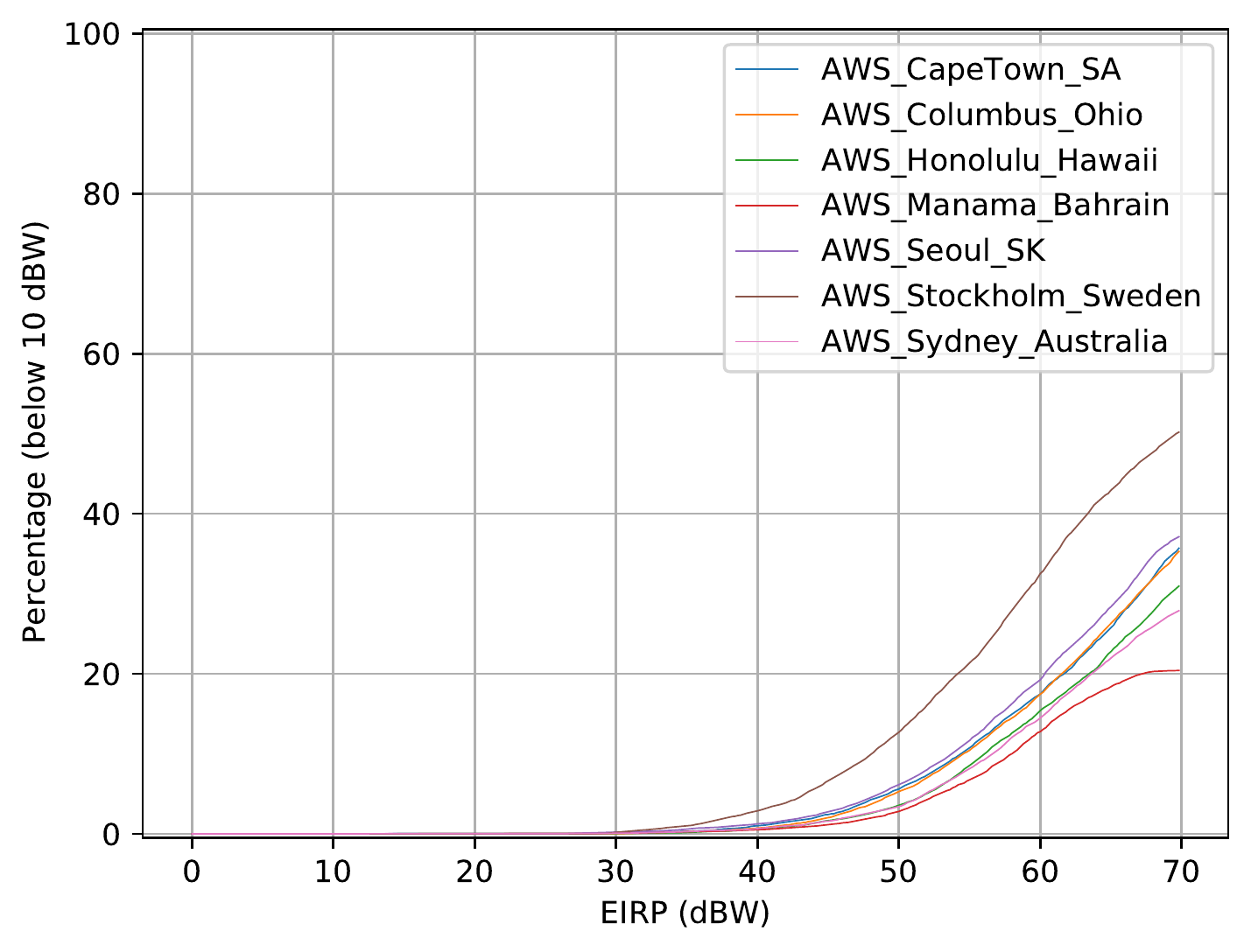}
            \caption[]{Planet Labs' Dove}
        \end{subfigure}%
        \begin{subfigure}{.5\textwidth}
            \includegraphics[scale=0.55]{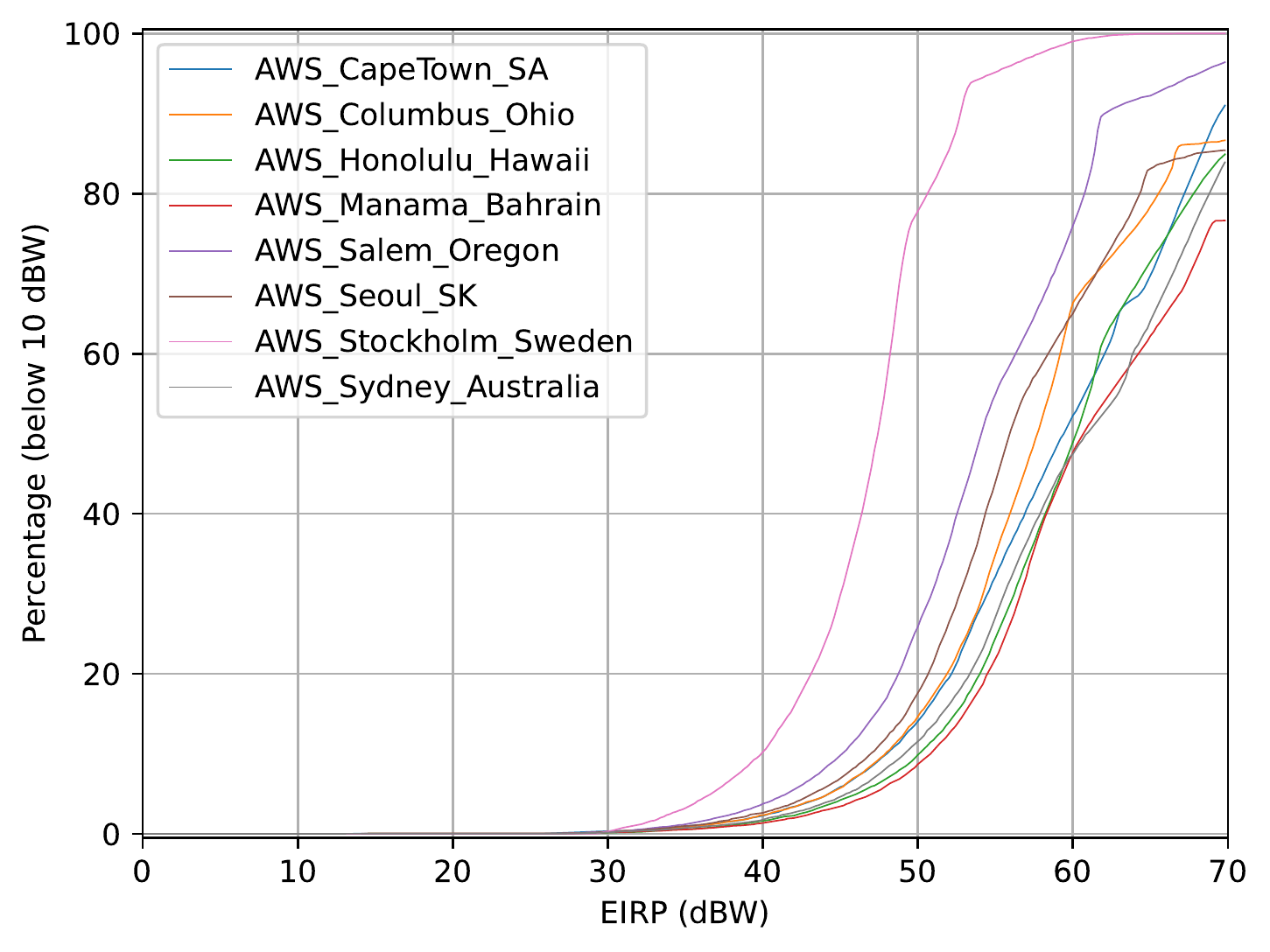}
            \caption[]{CubeSat}        
        \end{subfigure}%
    \caption{Groundstation curves for all constellations, showing percentage time jamming as a function of satellite power, with a threshold of 10 dBW.}
    \end{center}
\end{figure*}

\bibliographystyle{bib/IEEEtran}
\bibliography{bib/sample, bib/Similar}

\vspace{12pt}

\end{document}